\documentclass[12pt]{article}
\usepackage{amsfonts}
\usepackage{amssymb}
\usepackage{amsmath}
\usepackage{mathrsfs}
\usepackage{bbm}
\usepackage{graphicx}
\usepackage{subfigure}
\usepackage{color}
\usepackage{verbatim}
\usepackage{algorithm}
\usepackage{amsmath}
\usepackage{algpseudocode}
\usepackage{setspace}
\usepackage{geometry}             
\usepackage{epstopdf}
\usepackage{cases}
\usepackage{cite}

\definecolor{rossoCP3}{cmyk}{0,.88,.77,.40}

\usepackage[hyperindex=true,
          pdfstartview=FitH,
          bookmarksnumbered=true,
          bookmarksopen=true,
          citecolor=blue,
          linkcolor=blue,
          colorlinks=true,
          urlcolor=rossoCP3,
          unicode]{hyperref}

\begin{document}

\title{\bf
Ruppeiner Geometry of the RN-AdS Black Hole Using Shadow Formalism}
{\author{\small Chao Wang${}^{1,2}${}, Bin Wu${}^{1,2,3,4}$\thanks{{\em email}: \href{mailto:binwu@nwu.edu.cn}{binwu@nwu.edu.cn}}{ }, Zhen-Ming Xu${}^{1,2,3,4}${}, and Wen-Li Yang${}^{1,2,3,4}$
		\vspace{5pt}\\
		\small $^{1}${\it School of Physics, Northwest University, Xi'an 710127, China}\\
		\small $^{2}${\it Institute of Modern Physics, Northwest University, Xi'an 710127, China}\\
		\small $^{3}${\it Shaanxi Key Laboratory for Theoretical Physics Frontiers, Xi'an 710127, China}\\
		\small $^{4}${\it Peng Huanwu Center for Fundamental Theory, Xi'an 710127, China}
	}
	
\date{}
\maketitle

\begin{spacing}{1.2}
\begin{abstract}
The connection between the shadow radius and the Ruppeiner geometry of a charged static spherically symmetric black hole is investigated. The normalized curvature scalar is adopted, and its close relation to the Van der Waals-like and Hawking-Page phase transition of Reissner-Nordstr\"{o}m AdS black hole is studied. The results show that the shadow radius is a useful tool to reveal the correct information of the phase structure and the underlying microstructure of the black hole, which opens a new window to investigate the strong gravity system from the observational point of view.
\end{abstract}

\section{\textsf{Introduction}}

The first image of the black hole at the center of the galaxy M87* released by Event Horizon Telescope (EHT) \cite{Akiyama:2019cqa,Akiyama:2019bqs}, parallelly with the gravitational wave detection by LIGO and Virgo \cite{Abbott:2016blz}, have been considered as two great achievements of the Einstein gravity. It also ignites our enthusiasm for researching the shadow of the black hole which is the consequence of the interaction between the photon and the strong gravitational field. Precisely, not all the light can be detected by a static observer, some of them will drop into the black hole, which indicated that the shadow is closely connected to its dynamic. For black holes, we can exploit its shadow to obtain its information  \cite{Zakharov:2011zz,Tsukamoto:2014tja,Kumar:2018ple,Allahyari:2019jqz,Khodadi:2020jij}, such as the mass, the charge, and the rotation parameters, even the matter distribution around black holes \cite{Xu:2018mkl,Saurabh:2020zqg}. Numerous works have been done to investigate the properties of the black hole shadow. The dependence of the shadow of Schwarzschild black hole
on the location of the observer and black hole mass is studied in \cite{Synge:1966}. The shape of a rotating Kerr black hole shadow is affected by the dragging effect caused by rotation \cite{Vries:2000,Hioki:2008zw,Eiroa:2017uuq}. The properties of the shadow have been generalized to various types of black holes in \cite{Bambi:2008jg,Bambi:2010hf,Wang:2017qhh,Guo:2018kis,Yan:2019etp,Konoplya:2019sns}. More developments on black hole shadow can be found in a brief review \cite{Cunha:2018acu}.

After the seminal work by Hawking and Bekenstein about the temperature and entropy of black holes \cite{Bekenstein:1973ur,Hawking:1974sw,Bardeen:1973gs,Hawking:1976de}, the thermodynamics of black holes have attracted much attention from the researchers. By introducing the relation between the Hawking-Page phase transition \cite{Hawking:1982dh} and the confinement/deconfinement phase transition of the gauge field \cite{Witten:1998zw}, the phase structure of black holes in AdS spacetime becomes more charming. Interpreting the negative cosmological constant as the thermodynamic pressure \cite{Kastor:2009wy,Dolan:2011xt,Cvetic:2010jb} enriches the phase structure of the black hole, such as the small/large phase transition of the charged AdS black hole which is similar to the liquid/gas phase transition of the Van der Waals fluid \cite{Kubiznak:2012wp}. In line with that, the phase transition and critical behavior of the black hole in the extended phase space have been numerously investigated in \cite{Kubiznak:2016qmn,Toledo:2019amt,Hendi:2012um,Wei:2012ui,Cai:2013qga,Zhao:2013oza,Mo:2013ela,Altamirano:2013ane,Spallucci:2013osa,Xu:2014tja,Miao:2018fke,Miao:2016ulg,Xu:2013zea}. Under these efforts, black holes have been widely recognized as a thermodynamic system.
What’s more, the observation of black hole opens a new window for its thermodynamic research. As a strong gravitational system, the trajectory of the photon moving around the black hole has been proved to be an effective tool to reflect its phase structure \cite{Wei:2017mwc,Wei:2018aqm,Zhang:2019tzi,Zhang:2019glo,Belhaj:2020nqy}.

Despite the success of black hole thermodynamics, its microscopic origin is a mystery at present. The Ruppeiner geometry \cite{PhysRevA.24.488,Ruppeiner:1983zz,Ruppeiner:1995zz} derived from the laws of thermodynamic fluctuations gives us a glimpse of the microstructure of black holes. The Ruppeiner line element measures the distance between two adjacent thermodynamic fluctuation states, it is defined as
\begin{equation}
	\Delta l^2= - \frac{\partial^2 S}{\partial X^\mu \partial X^\nu} \Delta X^\mu X^\nu = g^R_{\mu \nu} \Delta X^\mu \Delta X^\nu,    \label{Ruppeiner}
\end{equation}
from which we can obtain the thermodynamic Riemannian curvature scalar, where $S$ is the entropy of the system and $X^\mu$ is thermodynamic coordinates depending on the chosen of the basic thermodynamic differential relation. The corresponding curvature scalar marks the interaction between adjacent modules of the fluid system: a positive (negative) thermodynamic curvature scalar indicates that a repulsive (attractive) interaction domain and a noninteracting system such as the ideal gas corresponds to the flat Ruppeiner metric \cite{Ruppeiner:2008kd,Ruppeiner:2010,Sahay:2010wi,Zhang:2015ova,Wei:2017icx,Chaturvedi:2017vgq,Xu:2019gqm,Wei:2019yvs,Xu:2020ftx}. The basic thermodynamic differential relation is usually chosen as $\mathrm{d} U= T \mathrm{d} S- P \mathrm{d} V$, which has the advantage to exploit the fine structure of Reissner-Nordstr\"{o}m AdS (RN-AdS) black hole thermodynamic phase more clearly \cite{Wei:2019uqg}. Alternatively, we can also set the basic thermodynamic differential relation as $\mathrm{d}H=T \mathrm{d} S+ V \mathrm{d} P$ \cite{Xu:2020gud}, which avoid the problem that the heat capacity at constant volume always vanish.

In the literatures \cite{Cai:2021fpr,Cai:2021uov}, the authors studied the relations between the shadow of the black hole and the thermodynamic curvature scalar with line element coming from the differential relation of enthalpy. In this paper, in order to reflect the thermodynamic fine phase structure of the black hole, we take the basic thermodynamic differential relation as $\mathrm{d} U= T \mathrm{d} S- P \mathrm{d} V$, and adopt the normalized Riemannian curvature scalar to study its connection to the shadow of the black hole. We study the small/large transition and the Hawking-Page transition using shadow analysis and expect that the observation of the shadow can directly reveal the thermodynamic phase structure and potential microstructure of the black hole.

The outline of this paper is as follows. In section \ref{I}, the shadow of RN-AdS black holes has been reviewed briefly. The Van der Waals-like phase transition and the underlying microstructure are studied in section \ref{II}. In section \ref{III}, we investigate the Hawking-Page phase transition of the black hole, and further explore its underlying microstructure. Throughout this paper, we apply the units $\hbar= k_B =c =G =1$.

\section{\textsf{The shadow radius of RN-AdS black hole}}\label{I}

The line element describing a static spherically symmetric black hole is expressed as 
\begin{equation}
	\mathrm{d} s^2 = -f(r) \mathrm{d} t^2 + \frac{ \mathrm{d} r^2 }{ \textsl{g} (r) }  + r^2 ( \mathrm{d} \theta^2 + \sin^2 \theta \mathrm{d} \phi ^2),    \nonumber
\end{equation}
where $f(r)$ and $\textsl{g}(r)$ are the functions of the radius parameter $r$.
The Hamiltonian of the photon with sphere orbit around the black hole is \cite{Cunha:2018acu}
\begin{equation}
	 \mathcal{H} = \frac{1}{2} g_{\mu \nu} P^\mu P^\nu=0,    \nonumber
\end{equation}
where $P^\mu$ is the proper 4-momentum of the photon. Without loss of generality, we consider the photons moving on the equatorial plane with $\theta=\pi/2$. Thus the Hamiltonian is cast to
\begin{equation}
	\frac{1}{2}\left[-\frac{P_{t}^{2}}{f(r)}+ \textsl{g}(r) P_{r}^{2}+\frac{P_{\phi}^{2}}{r^{2}}\right]=0.    \label{Hamiltonian}
\end{equation}
Since $t$ and $\phi$ are not obviously included in the Hamiltonian, so that $\partial_t$ and $\partial_\phi$ are killing vectors, and the corresponding conserved constants are
\begin{equation}
	E=-P_t =-\frac{\partial \mathcal{H}}{\partial \dot{t}} \quad \mathrm{and} \quad L=P_\phi= \frac{\partial \mathcal{H}}{ \partial \dot{\phi}} ,
\end{equation} 
where $E$ and $L$ represent the energy and angular momentum of the photon, respectively. 
And the equation of motion derived from the Hamiltonian Eq.(\ref{Hamiltonian}) is written as
\begin{equation}
		\dot{t}=\frac{\partial \mathcal{H}}{\partial P_{t}}=-\frac{P_{t}}{f(r)}, \quad
		\dot{\phi}=\frac{\partial \mathcal{H}}{\partial P_{\phi}}=\frac{P_{\phi}}{r^{2}}, \quad
		\dot{r}=\frac{\partial \mathcal{H}}{\partial P_{r}}=P_{r} \textsl{g}(r),    \label{motion1}
\end{equation}
in which the dot denotes the derivative with respect to the affine parameter. The parameterized orbit of the photon deduced from Eq.(\ref{motion1}) has the form of
\begin{equation}
	\frac{\mathrm{d} r}{\mathrm{d}\phi} = \frac{\dot{r}}{\dot{\phi}} = \frac{r^2 \textsl{g}(r) P_r}{L}.   \label{motion2}
\end{equation}
The effective potential of the photon can be split from the Hamiltonian
\begin{equation}
	V_\mathrm{eff}(r) + \dot{r}^2 =0,    \nonumber
\end{equation}
which is given from Eq.(\ref{Hamiltonian}) and Eq.(\ref{motion1}) as the form of
\begin{equation}
	V_\mathrm{eff}(r)=\textsl{g}(r)\left[\frac{L^{2}}{r^{2}}-\frac{E^{2}}{f(r)}\right].    \label{veff}
\end{equation}
The orbit of the photon moving around the black hole is deduced straightly from the conditions that
\begin{equation}
	V_{\mathrm{eff}}(r)=V'_{\mathrm{eff}}(r)=0, \quad V''_{\mathrm{eff}}(r) > 0.  \label{null}
\end{equation}
Here the prime represent the derivate with respect to the radius of spacetime. Substituting Eq.(\ref{veff}) into Eq.(\ref{null}), one can easily obtain the relation that
\begin{equation}
	\frac{L}{E} = \frac{r}{\sqrt{f(r)}} \bigg|_{r_\mathrm{p}},  \label{up}
\end{equation}
in which $r_{\mathrm{p}}$ is the radius of the photon sphere around the black hole.
From Eq.(\ref{Hamiltonian}), Eq.(\ref{motion2}) and Eq.(\ref{up}), the parameterized orbit of the photon turns to
\begin{equation}
	\frac{\mathrm{d} r}{\mathrm{d} \phi} = 
	         \pm r \sqrt{\textsl{g}(r) \left( \frac{r^2 E^2}{f(r) L^2} -1 \right)}.    \label{turn}
\end{equation}
The perigee of null geodesic is characterized by the mathematical constraint $\frac{\mathrm{d} r}{\mathrm{d} \phi}\big|_{R}=0$, with which the photon sphere Eq.(\ref{turn}) reduces to
\begin{equation}
	\frac{\mathrm{d} r}{\mathrm{d} \phi} = \pm r \sqrt{\textsl{g}(r) \left( \frac{r^2 f(R)}{f(r) R^2} -1 \right)}.    \label{turnning}
\end{equation}
In Fig.\ref{Fig1}, we illustrate the black hole shadow detected by the observer located at $r_o$ with an angular $\alpha$, it yields
\begin{equation}
	\cot{\alpha} = \sqrt{ \frac{g_{rr}}{g_{\phi \phi}} }  \frac{\mathrm{d} r}{\mathrm{d} \phi}  \bigg|_{r=r_o}
	    =\frac{1}{r \sqrt{\textsl{g}(r)}} \frac{\mathrm{d} r}{\mathrm{d} \phi}  \bigg|_{r=r_o}.    \label{alpha}
\end{equation}
\begin{figure}[!h]
	\centering
	\includegraphics[width=15cm]{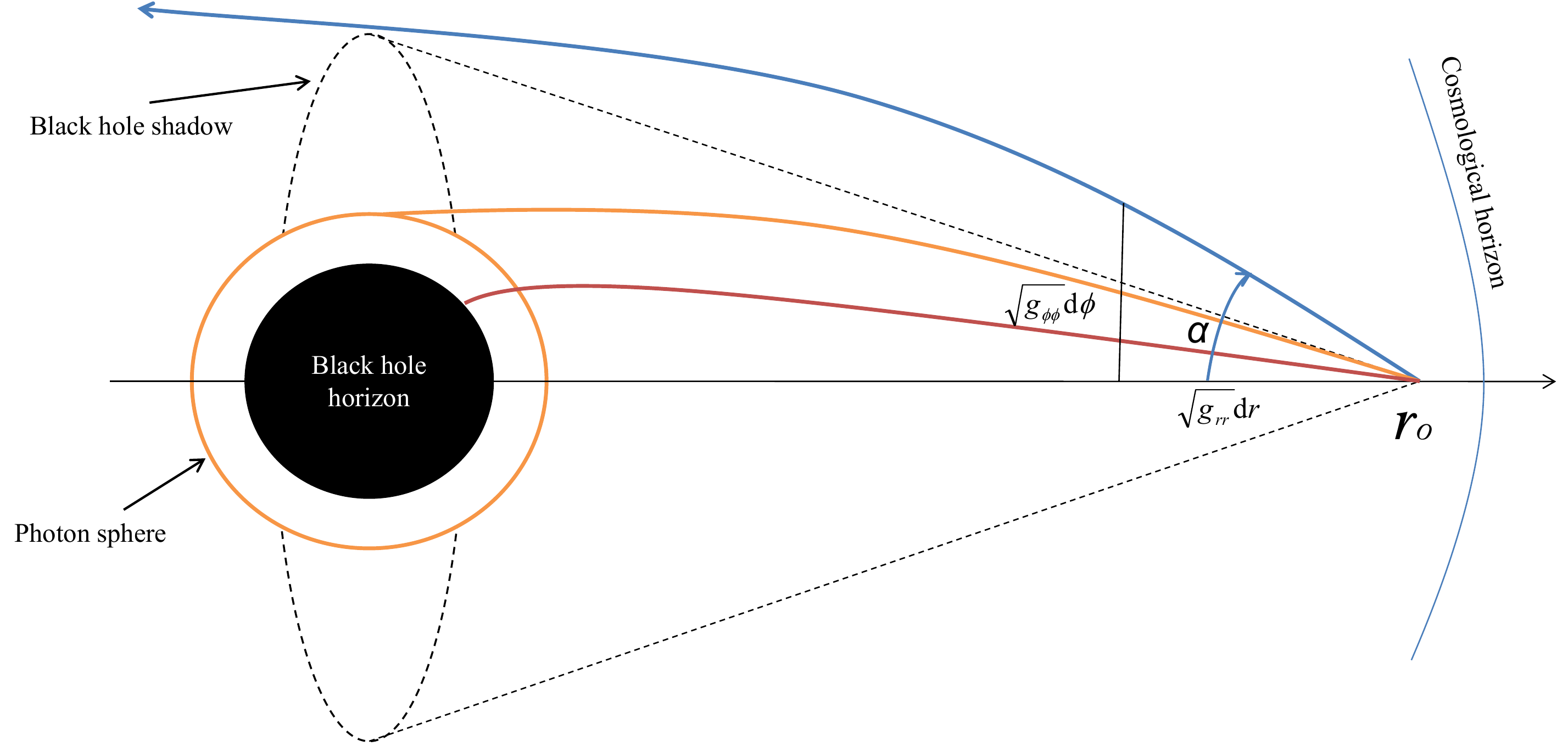}
	\caption{The RN-AdS black hole shadow observed by an observer at $r_o$.}\label{Fig1}
\end{figure}
Substituting Eq.(\ref{turnning}) into Eq.(\ref{alpha}), with the elementary trigonometry, we get the radius of the shadow for a spherically symmetric black hole in the sky of the observer at $r_o$ as
\begin{equation}
	r_s =r_o \tan{\alpha} \approx  r_o \sin{\alpha} = R \sqrt{\frac{f(r_o)}{f(R)}} \bigg|_{R\rightarrow r_\mathrm{p}},    \nonumber
\end{equation}
for small value of $\alpha$\footnote{The small $\alpha$ limit is just to simplify the calculation, in fact it won’t have much impact on the qualitative analysis without this approximation.}.

As two important subjects of black hole research, the character quantities of phase transition and the Ruppeiner geometry are the specific heat capacity $C$ and Ruppeiner curvature scalar $R$. With the thermodynamic entropy of black hole $S= \pi r^2_h$, we have
\begin{align}
	&\operatorname{Sgn}(C)=\operatorname{Sgn}\left(\frac{\partial T}{\partial r_{h}}\right)=\operatorname{Sgn}\left(\frac{\partial T}{\partial r_{s}} \frac{\mathrm{d} r_{s}}{\mathrm{d} r_{h}}\right),    \label{C}\\
	&\operatorname{Sgn}\left(\frac{\partial R}{\partial S}\right)=\operatorname{Sgn}\left(\frac{\partial R}{\partial r_s} \frac{\mathrm{d} r_s}{\mathrm{d} r_h} \frac{\mathrm{d} r_h}{ \mathrm{d} S}\right).    \label{R}
\end{align}
The equations above indicate that $r_h$ can be replaced by $r_s$ as a crucial parameter to directly reflect the thermodynamic phase structure and underlying microstructure of the black hole when $\mathrm{d}r_h / \mathrm{d} r_s$ is positive. 

In what follows, we take the RN-AdS black hole as a concrete example, whose metric has the form of
\begin{equation}
	f(r)=\textsl{g}(r) = 1- \frac{2M}{r} + \frac{ Q^2 }{ r^2 } + \frac{ 8 \pi P r^2 }{ 3 }.    \nonumber
\end{equation}
The parameters $M$ and $Q$ represent the black hole mass and charge, respectively. $P$ is the thermadynamic pressure related to the cosmological constant by $P=- \frac{\Lambda}{8 \pi }$ \cite{Kastor:2009wy,Dolan:2011xt,Cvetic:2010jb}. In terms of pressure $P$ and the event horizen radius $r_h$, the mass $M$ is expressed as
\begin{equation}
	M = \frac{r_h}{2} + \frac{Q \Phi}{2} + \frac{4 \pi P r^3_h }{3},    \label{mass}
\end{equation}
in which $\Phi=Q/r_h$ is the electrical potential. The Hawking temperature of the RN-AdS black hole is 
\begin{align}
	T = \frac{1}{ 4 \pi r_h } \left( 1+ 8 \pi P r^2 _h - \frac{ Q^2 }{ r^2_h } \right).    \label{temperature}
\end{align}
The thermodynamic volume conjugated to the pressure arrives at
\begin{equation}
	V=\left( \frac{\partial M}{\partial P}\right)_{Q,S}=\frac{4}{3} \pi r^3_h= \frac{\pi}{6} v^3,    \label{volume}
\end{equation}
here $v=2r_h$ denotes the ``specific volume''. With the Hawking temperature Eq.(\ref{temperature}) and thermodynamic volume Eq.(\ref{volume}), we get the state equation as follow
\begin{equation}
	P=\frac{T}{v}-\frac{1}{2 \pi v^2} +\frac{2Q^2}{\pi v^4}.    \label{state1}
\end{equation}
Analogy to the Van der Waals fluid, the black hole undergoes a second-order phase transition, whose critical point can be obtained under the conditions $(\partial_v P)_T=(\partial_{v,v} P)_T=0$ and $(\partial_{v,v} P)_T < 0$ that
\begin{equation}
	P_{c}=\frac{1}{96 \pi Q^2}, \quad r_{h_{c}}=\sqrt{6} Q, \quad \Phi_c=\frac{1}{\sqrt{6}}, \quad T_{c}=\frac{1}{3 \sqrt{6} \pi Q} \quad \text{and} \quad S_c=6\pi Q^2.    \label{critical}
\end{equation}
In the reduced parameter space, the equation of state turns to
\begin{equation}
	\tilde{P}=\frac{8 \tilde{T}}{3 \tilde{v}}-\frac{2}{\tilde{v}^2}+\frac{1}{3 \tilde{v}^4},    \label{state2}
\end{equation}
where the dimensionless reduced parameters are defined as
\begin{equation}
	\tilde{P}= \frac{P}{P_c}, \quad \tilde{r}_h=\frac{r_h}{r_{hc}} \quad \text { and } \quad \tilde{T}=\frac{T}{T_c}.    \label{reduced}
\end{equation}

To obtain the shadow radius of RN-AdS black hole, we should get the radius of photon sphere around the black hole. With the constraint Eq.(\ref{null}), we obtain the radius of the photon sphere around the black hole as
\begin{equation}
	r_\mathrm{P}= \frac{1}{2} \left(3M+ \sqrt{9M^2 -8Q^2} \right).    \label{rp}
\end{equation}
The critical point Eq.(\ref{critical}) leads a critical quantity $r_\mathrm{P}= (2+\sqrt{6})Q$, which turns the shadow radius to 
\begin{equation}
	\tilde{r}_s= \frac{\tilde{r}_\mathrm{P} \sqrt{f(\tilde{r}_o)}}{ 
		\sqrt{f(\tilde{r}_{\mathrm{P}})}}.    \label{rs}
\end{equation} 
Here $\tilde{r}_{\mathrm{P}}=r_\mathrm{P}/ r_{\mathrm{P}c}$, $\tilde{r}_o = r_o/ r_{\mathrm{P}c}$ and $\tilde{r}_s = r_s /r_{\mathrm{P}c}$ are the reduced parmeters.  In the following, all the quantities we used are dimensionless reduced parameters, so that we will omit the symbol `$\sim$' on the reduced parameters for simplification.

In order to check whether the shadow radius $r_s$ can be treated as the crucial parameter to directly reflect the thermodynamic phase structure and underlying microstructure of the black hole, we turn our attention to the slope of $\mathrm{d} r_s / \mathrm{d} r_h$.
\begin{figure}[!h]
	\centering
	{
		\includegraphics[width=9cm]{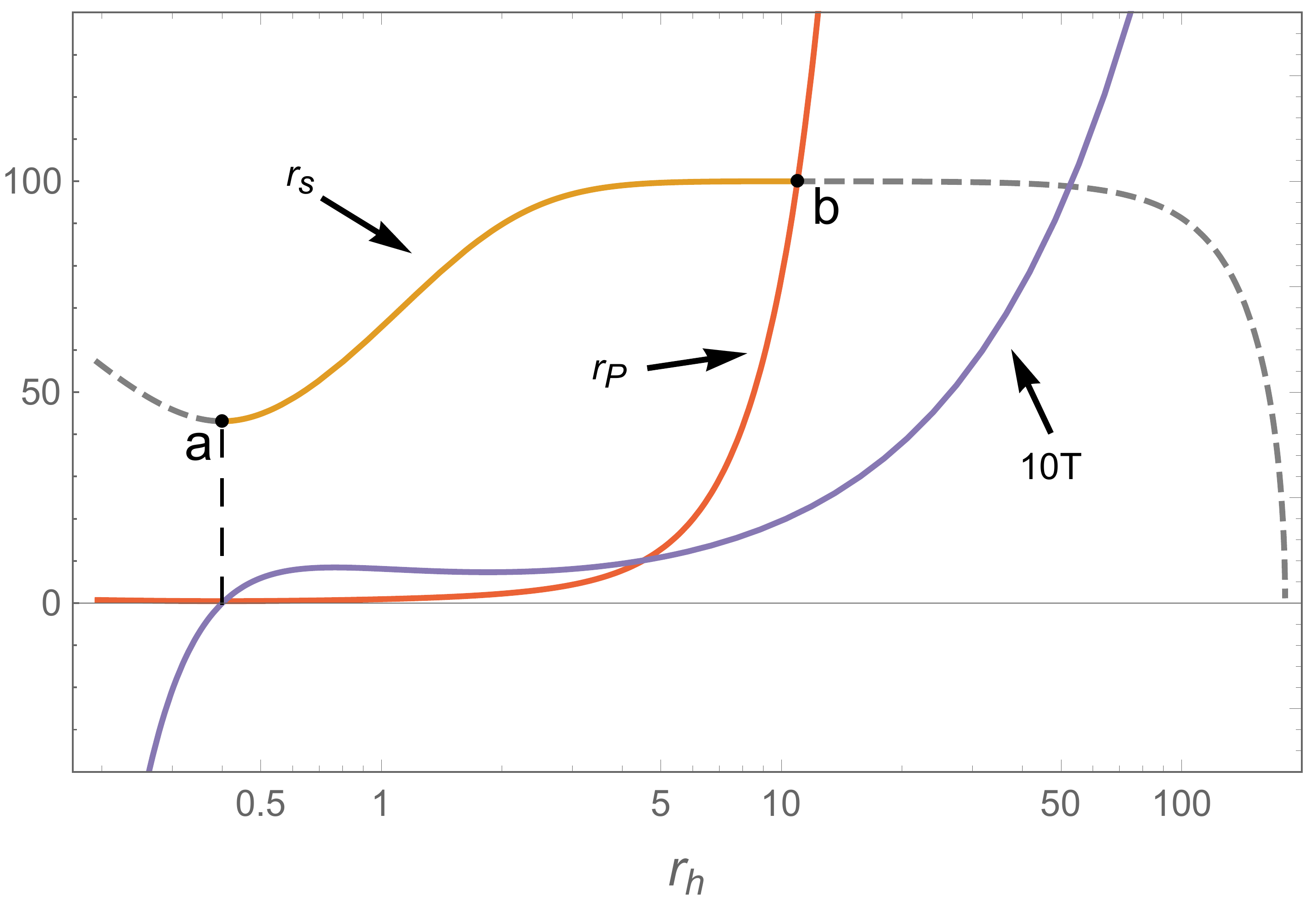}
	}
	\caption{The specific schematic diagram of the shadow seen by the observer located at $r_o=100$. Here we set $P=0.5$.}\label{Fig2}
\end{figure}  
In figure Fig.\ref{Fig2}, we plot the specific schematic diagram of the shadow seen by the observer located at $r_o=100$. As shown in the figure, these three solid curves with different colors represent the shadow radius, photon sphere radius and $10$ times of the reduced Hawking temperature, respectively. We can observe that there are two local extremum points marked as $a$ and $b$ in the graph, which divides the shadow curve into three parts. Only the part between these two points is competent to reflect the phase transition and the underlying microstructure of the black hole. And the remaining two segments are non-physical: the left part of $a$ corresponds to negative temperature, and the shadow radius larger than that of point $b$ is no longer applicable for the observer is always assumed to be located out of the photon sphere. As discussed above, the shadow radius can be a crucial parameter to reflect the phase structure and the underlying microstructure of the RN-AdS black hole directly.

\section{\textsf{VdW-like phase transition and microstructure}}\label{II}

In this section, we would like to study the Van der Waals-like phase transition and the underlying microstructure of RN-AdS black hole in four-dimensional spacetime from the perspective of observation. With the Hawking temperature Eq.(\ref{temperature}) and the shadow radius Eq.(\ref{rs}), we depict the isobar curves in Fig.\ref{Fig3} in reduced $(T,r_h)$ and $(T,r_s)$ space with reduced pressure $P=0.5$, $1$ and $1.5$ from bottom to top. Here and after we always set $r_o=100$.
\begin{figure}[!h]
	\centering
	\subfigure[]{
		\includegraphics[width=6cm]{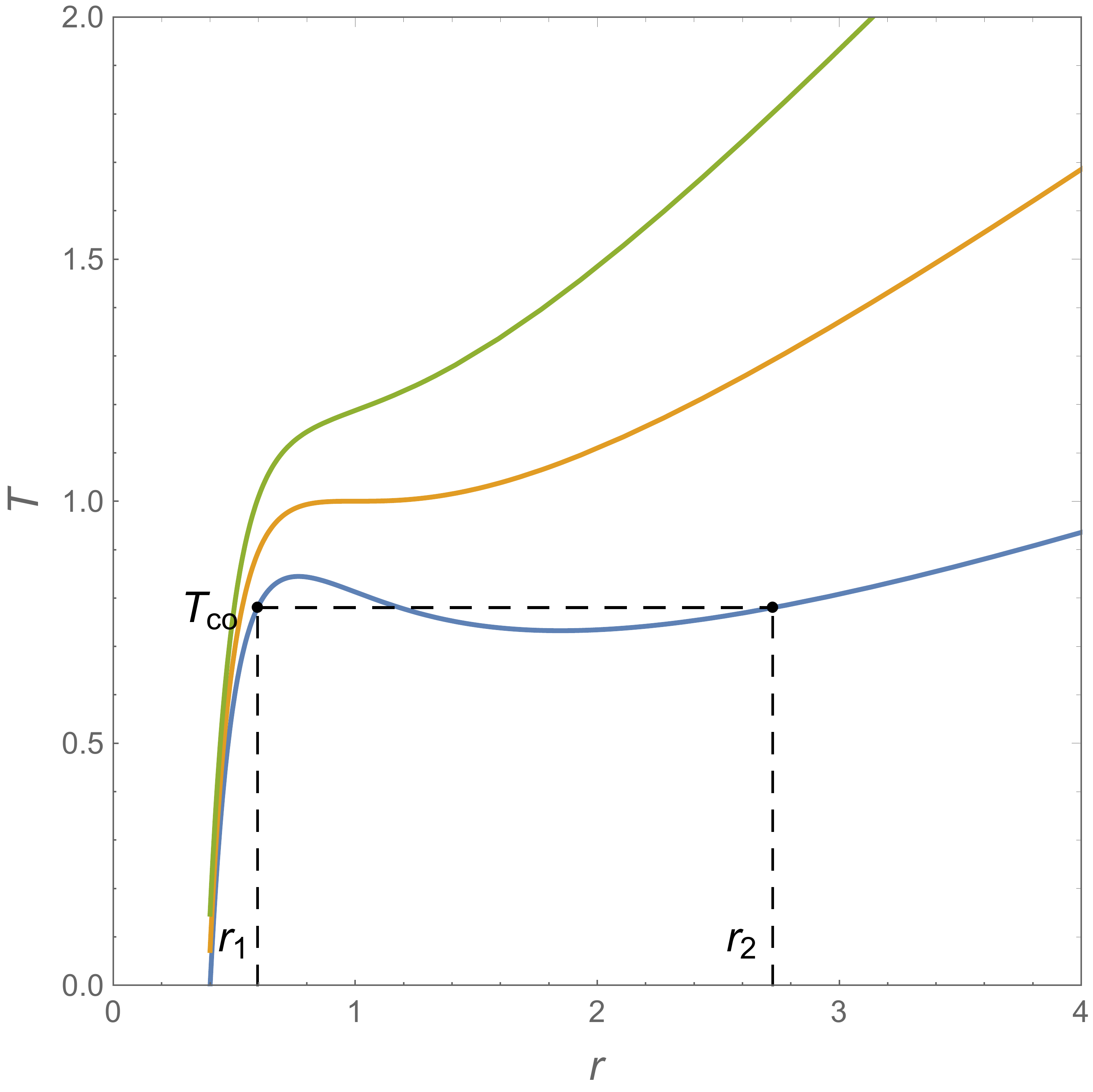}
	}
	\quad
	\subfigure[]{
		\includegraphics[width=6cm]{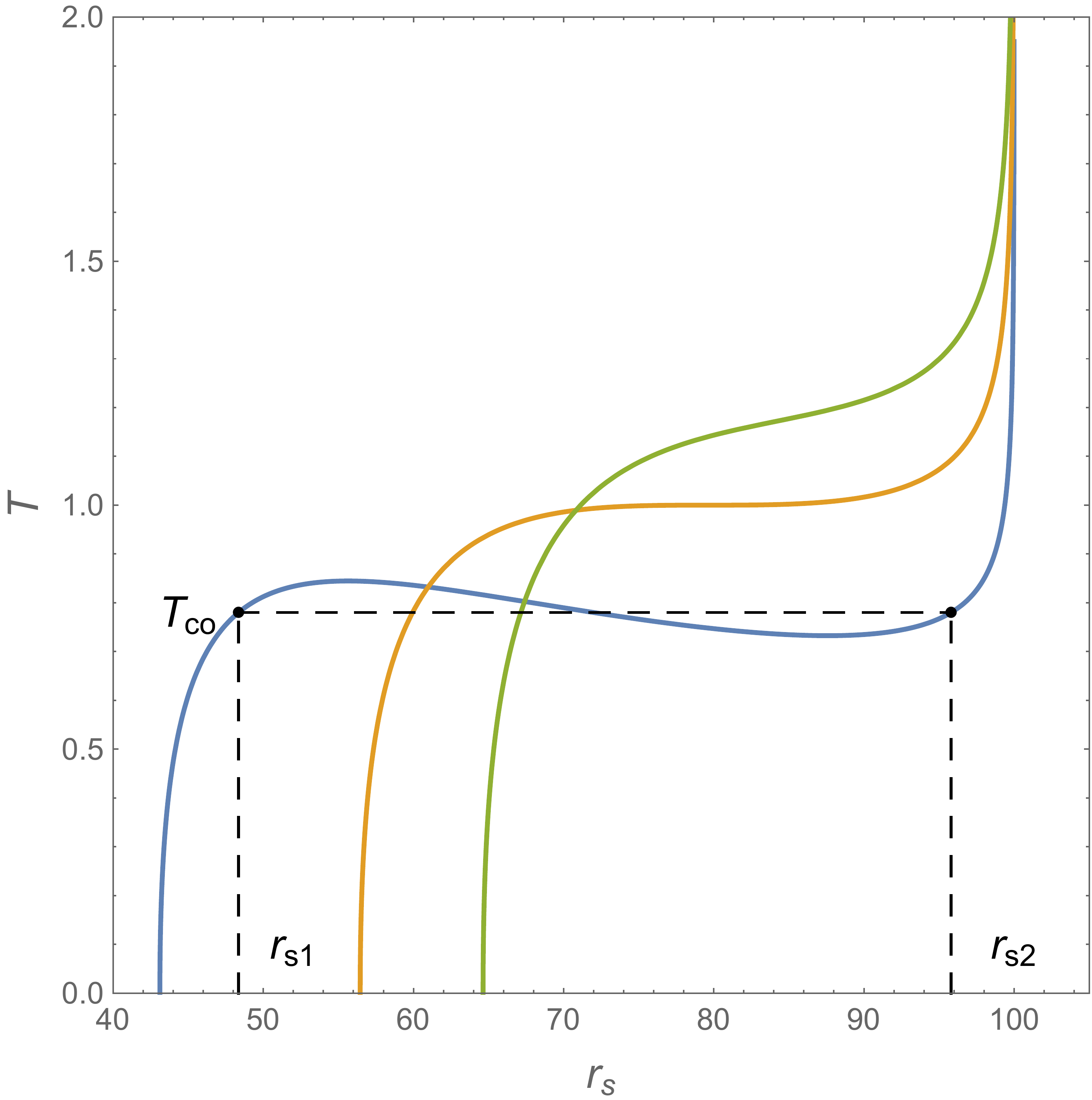}
	}
	\caption{Isobaric curve with $P=0.5$, $1$ and $1.5$ from bottom to top. $T_\mathrm{co}$ is the coexistence temperature.}\label{Fig3}
\end{figure}
It is clear that for $P > 1$, the temperature shows monotonous behavior in both figures which indicates that the black hole is in the supercritical phase. For small pressure $P< 1$, there are two local extreme points that divide the isobar curve into three parts. The part with positive slopes on both sides corresponds to the stable small/large black hole. While the rest segment corresponds to the unstable phase.  $T_\mathrm{co}$ is the coexistence temperature comes from the Maxwell equal area law in $(T, S)$ space, and the black hole phase is divided into three segments: the stable small black hole with horizon (shadow) smaller than $r_1$ ($r_{s1}$), while great than $r_2$ ($r_{s2}$) correspond to the stable large black hole. Through the above analysis, we see that shadow radius $r_s$ can indeed reflect the phase structure of the black holes. But you will still find some differences in Fig.\ref{Fig3}(b) by further compare these two phase diagrams. On one hand, there is a cut-off for the isobar curve, which is consistent with the results shown in Fig.\ref{Fig2} since the limitation on the position of the observer. On the other hand, there exist intersections between different isobars. This does not conflict with our understanding of the equation of state, because the intersection point corresponds to different thermodynamic volumes on different curves. 

To obtain the information of the underlying microstructure from the observation, we investigate the relationship between Ruppeiner thermodynamic geometry of RN-AdS black hole with shadow formalism. We chose the thermodynamic differential relation of internal energy $\mathrm{d} U =T \mathrm{d} S - P \mathrm{d} V$ as the basic one to build the thermodynamic line element in $(T, V)$ phase space which can be expressed as follows
\begin{equation}
	\Delta l^2=\frac{C_V}{T^2} \Delta T^2-\frac{(\partial_V P)_T}{T} \Delta V^2,    \label{IM}
\end{equation}
where $C_V=T(\partial_T S)_V$ is the heat capacity at constant volume, and the thermodynamic curvature scalar of the line element above is obtained by adopting the convention in \cite{Ruppeiner:1995zz}
\begin{align}
	R&=\frac{1}{2C_V^2(\partial_V P)^2}\bigg \{ T(\partial_V P) \Big[ (\partial_T C_V)(\partial_V P-T\partial_{T,V} P) + (\partial_V C_V)^2	\Big]  \nonumber \\
	& +C_V \Big[ (\partial_V P)^2+ T\left((\partial_V C_V)(\partial^2_V P)-T(\partial_{T,V} P)^2\right)+2T(\partial_V P)\left(T(\partial_{T,T,V} P)-(\partial^2_V C_V)\right)
	\Big] \bigg \}.    \label{sc1}
\end{align}
The heat capacity at constant volume always vanish for RN-AdS black hole, and that will cause the singularity of the thermodynamic metric. Therefore we introduce a normalized curvature scalar $R_N=R C_V$ as in Ref.\cite{Wei:2019uqg}, that is
\begin{equation}
	R_N=R C_V=\frac{\left(\partial_V P\right)^2-T^2\left(\partial_{V,T}P\right)^2+2T^2\left(\partial_V P\right)\left(\partial_{V,T,T}P\right)}
	{2\left(\partial_V P\right)^2}.    \label{sc2}
\end{equation}
Substituting the state equation Eq.(\ref{state1}) into Eq.(\ref{sc2}), the normalized
 curvature scalar of RN-AdS black hole with the reduced parameters arrives at
\begin{equation}
	R_N=\frac{\left(3V^\frac{2}{3}-1\right) \left(-4TV +3V^\frac{2}{3}-1\right) }
	         {2\left(-2TV +3V^\frac{2}{3}-1\right)^2}.    \label{sc3}
\end{equation}
To obtain the shadow boundary observed at $r_o=100$, we apply the stereographic projection as the function of the celestial coordinate $(x,y)$ \cite{Eiroa:2017uuq}
\begin{align}
	x &= \lim_{r \to \infty} -r^2 \sin \theta_0 \frac{\mathrm{d} \phi}{\mathrm{d} r}
	\bigg|_{\theta_0=\frac{\pi}{2}},    \label{x}\\
	y &= \lim_{r \to \infty} r^2 \frac{\mathrm{d} \theta}{\mathrm{d} r}.    \label{y}
\end{align}

By connecting Eq.(\ref{rs}) and Eq.(\ref{sc3}), we depict variation of $R_N$ on the silhouette of black hole shadow below. The shape of shadow comes from Eq.(\ref{x}) and Eq.(\ref{y}). In these shadow silhouette, the region marked as yellow surrounded by the black circle corresponds to the positive value of $R_N$.
\begin{itemize}
	\item $P =0.5$ in Fig.\ref{Fig6}:
	          \begin{figure}[!h]
	          	\centering
	          	\subfigure[]{
	          		\includegraphics[width=7.67cm]{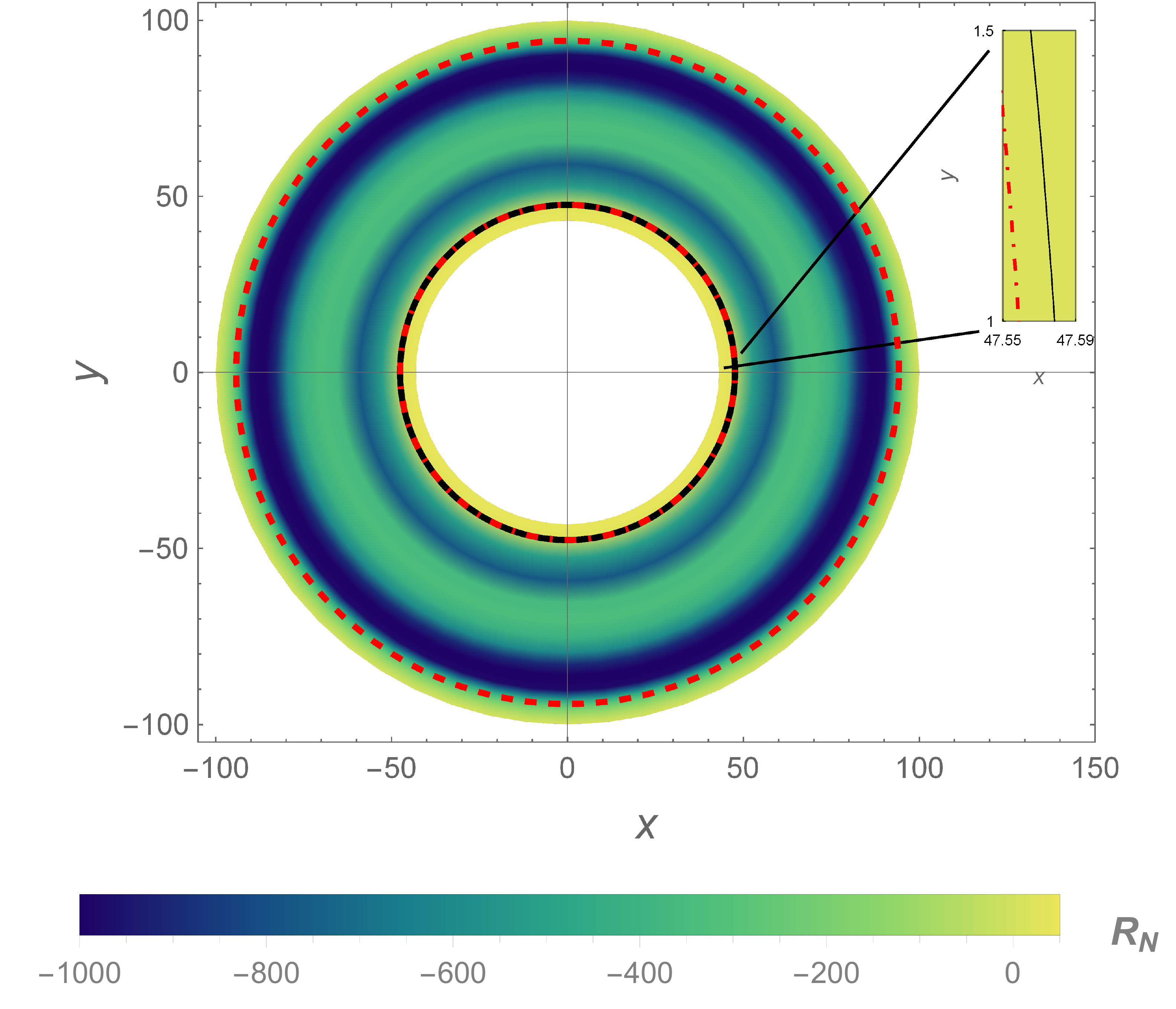}
	          	}
	          	\quad
	          	\subfigure[]{
	          		\includegraphics[width=6.8cm]{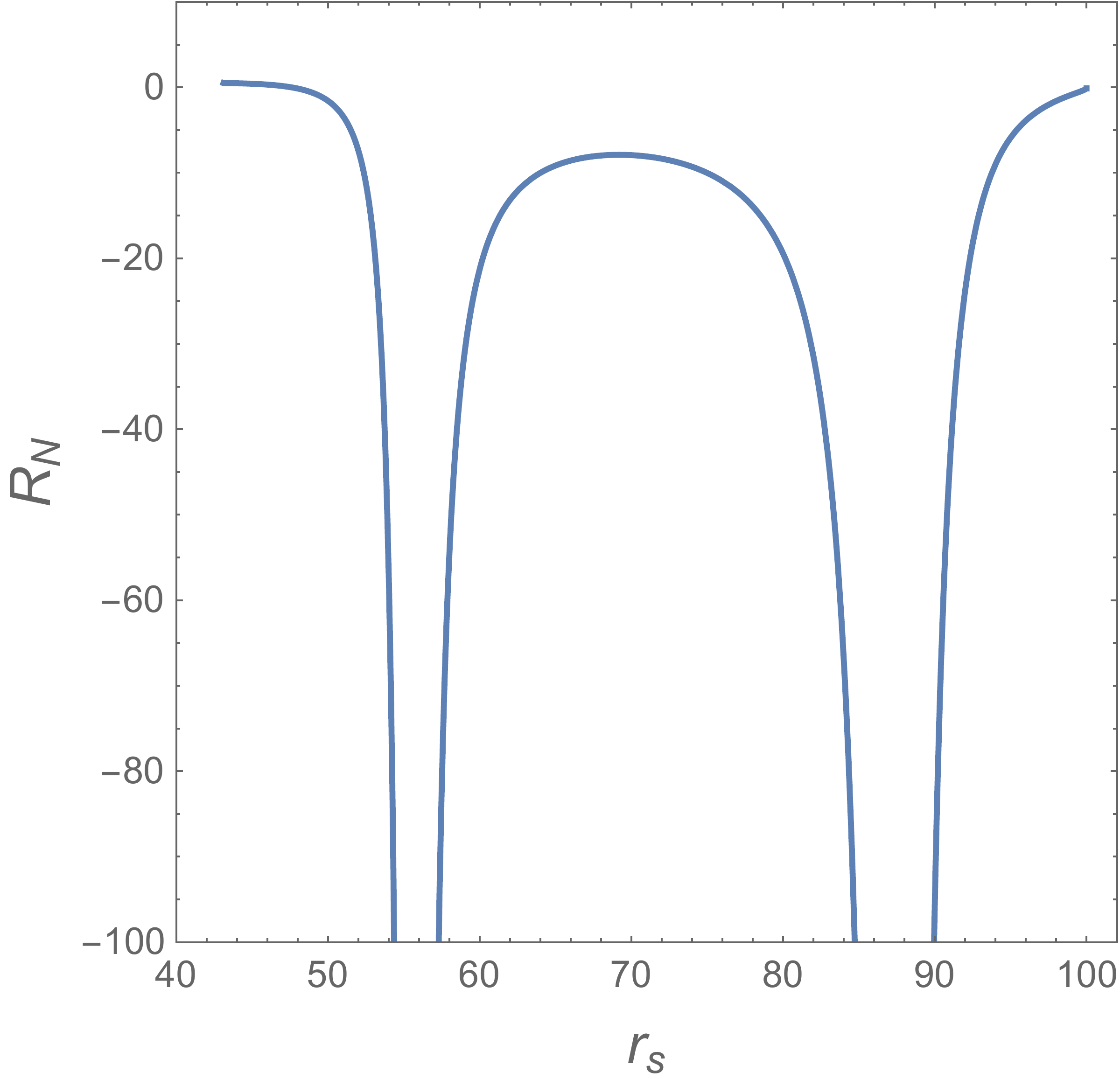}
	          	}
	          	\caption{The variation of $R_N$ in terms of the shadow radius $r_s$ with $P = 0.5$ and $r_o=100$. }\label{Fig6}
	          \end{figure}
	          The variation of the normalized curvature scalar with $P=0.5$ and $r_o=100$. We show the gradient with color, and the corresponding value is provided by the legend-bar, which show us that the value of $R_N$ of the black hole is smaller with deeper color. The dot-dashed and dashed red circle corresponding to $r_{s1}$ and $r_{s2}$ in Fig.\ref{Fig3}(b) represent the saturated small and large black hole, respectively. The epilogue in Fig.\ref{Fig6}(a) states that the interaction between the potential ``molecules'' of the saturated small black hole is repulsive domain. The shadow of black holes smaller than the red dot-dashed circle is the stable small black hole while larger than the red dashed circle corresponds to the stable large black hole phase. The black hole between these two red circles is divided into three phases by two circles: the metastable small, unstable, and metastable large phases. The deep blue region represents the divergence of the curvature scalar.
	\item $P = 1$ in Fig.\ref{Fig7}:
	          \begin{figure}[!h]
	          	\centering
	          	\subfigure[]{
	          		\includegraphics[width=6.8cm]{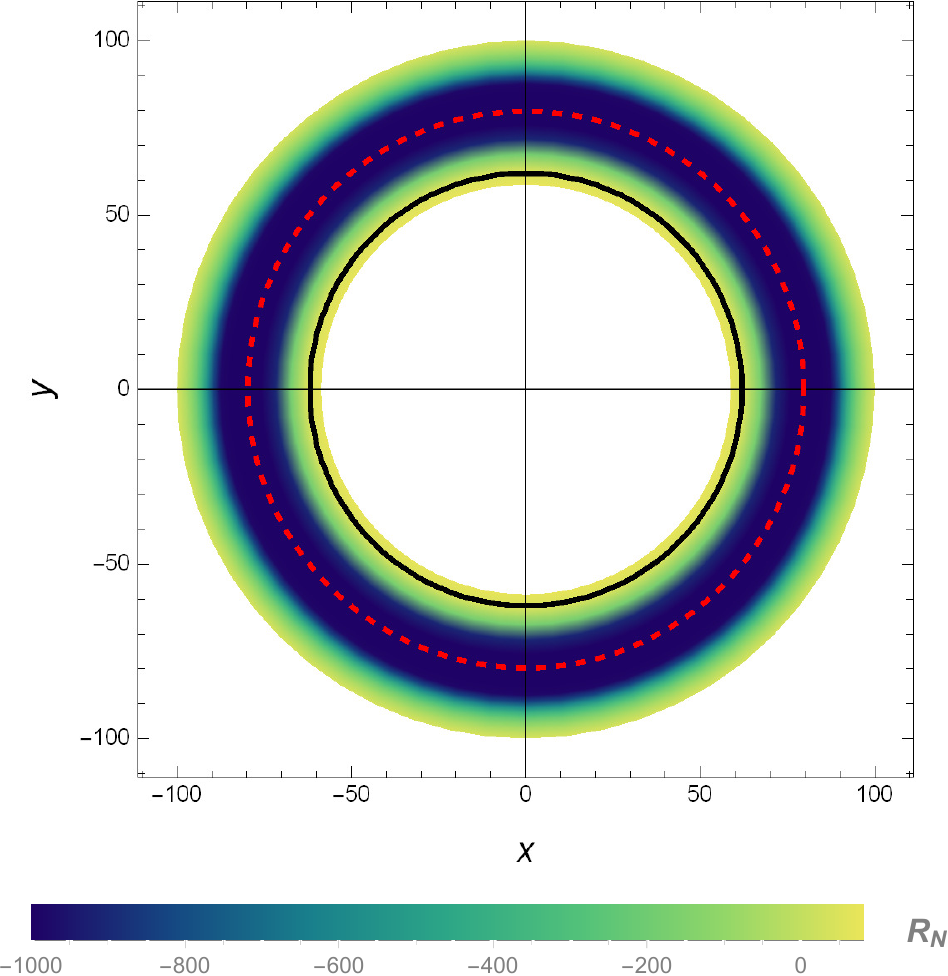}
	          	}
	          	\quad
	          	\subfigure[]{
	          		\includegraphics[width=7.5cm]{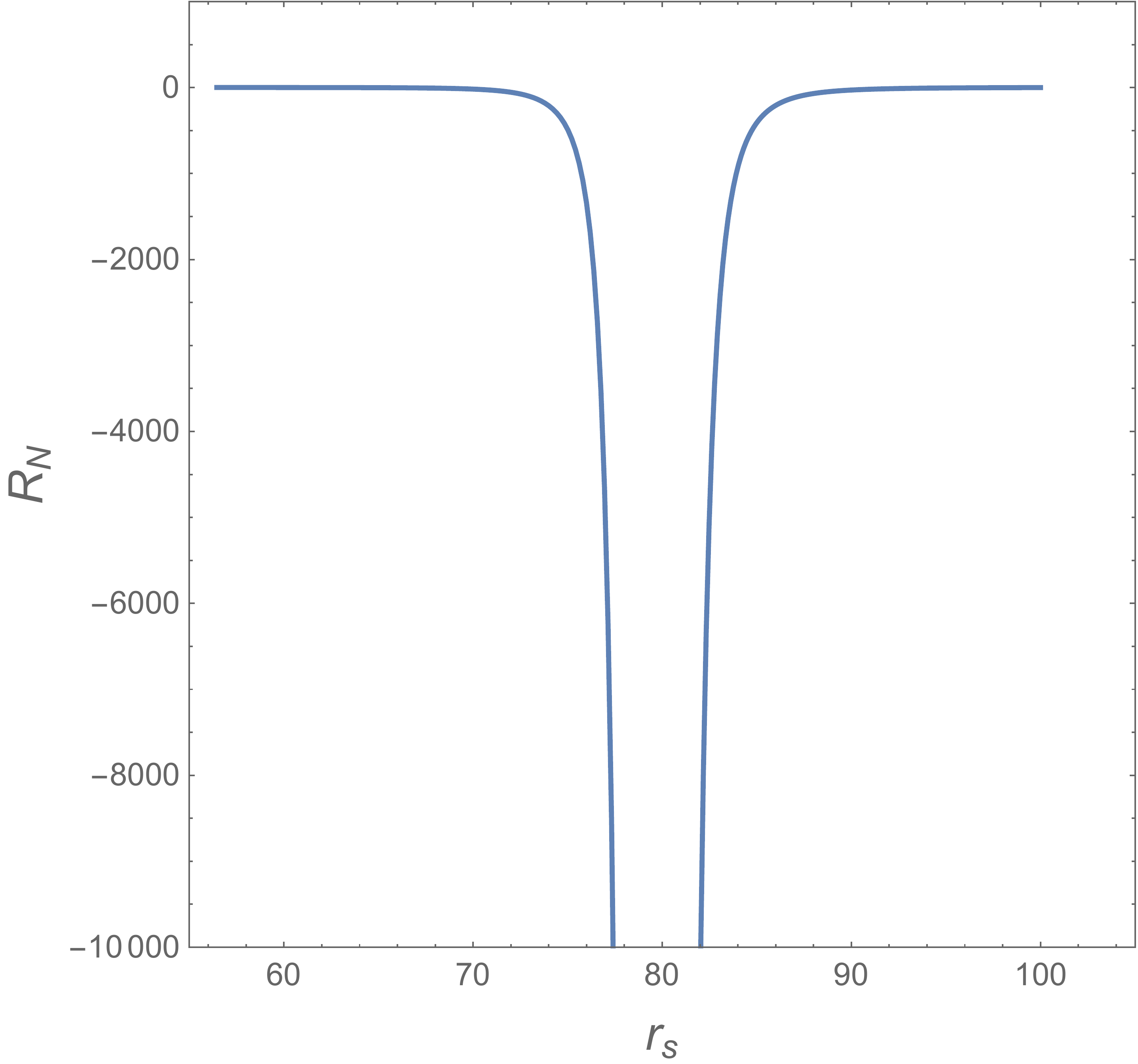}
	          	}
	          	\caption{The variation of $R_N$ in terms of the shadow radius $r_s$ with $P = 1$ and $r_o=100$. }\label{Fig7}
	          \end{figure}
	          The red dashed circle is the divergent point that corresponds to the critical point of the small/large black hole phase transition. At this time, the phase structure of the black hole is relatively simple. A black hole with a shadow radius smaller than the red dashed ring is the stable small black hole phase, and the rest part represents the stable large black hole phase.
	           \begin{figure}[!h]
	          	\centering
	          	\subfigure[]{
	          		\includegraphics[width=6.9cm]{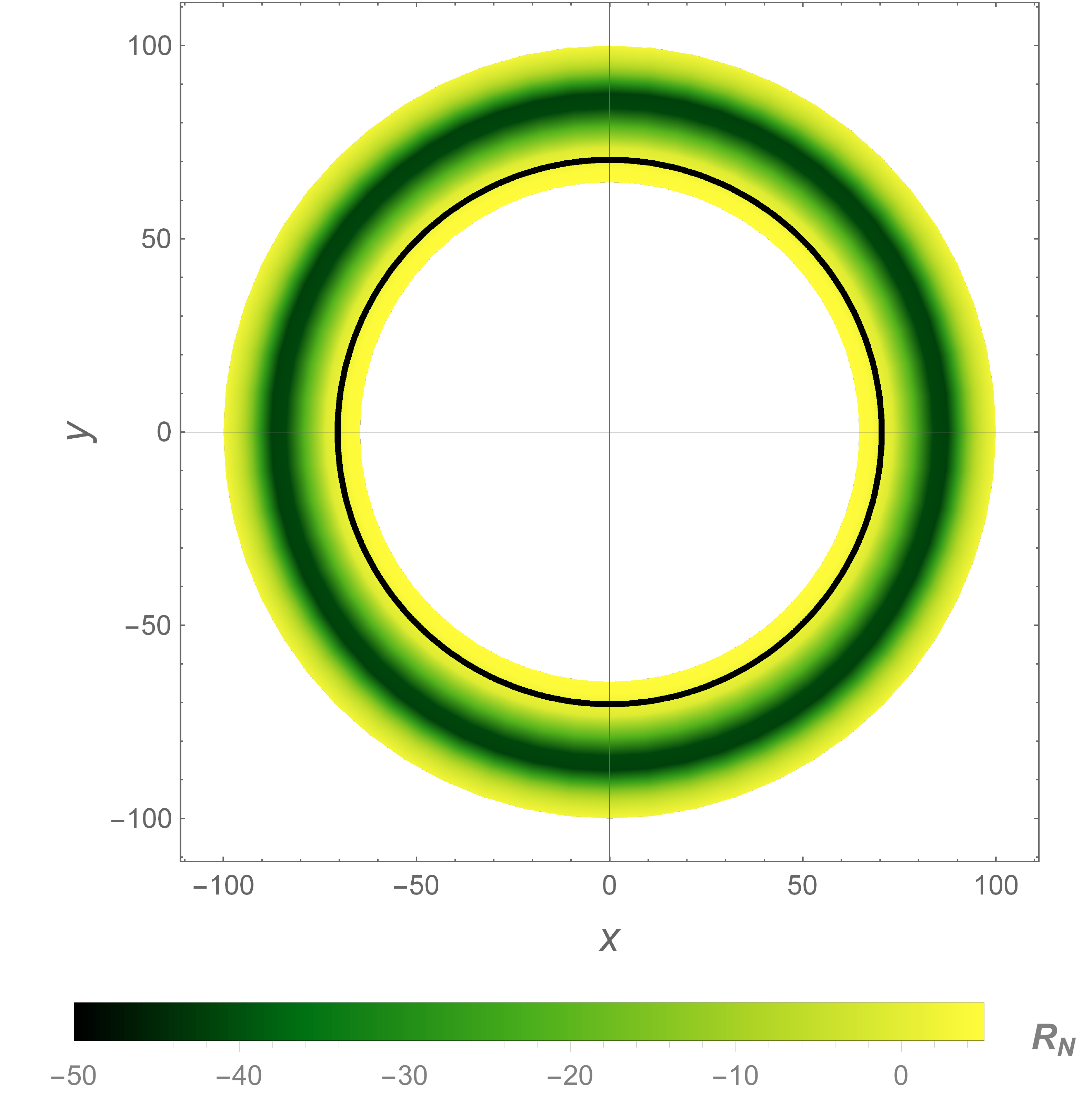}
	          	}
	          	\quad
	          	\subfigure[]{
	          		\includegraphics[width=7.1cm]{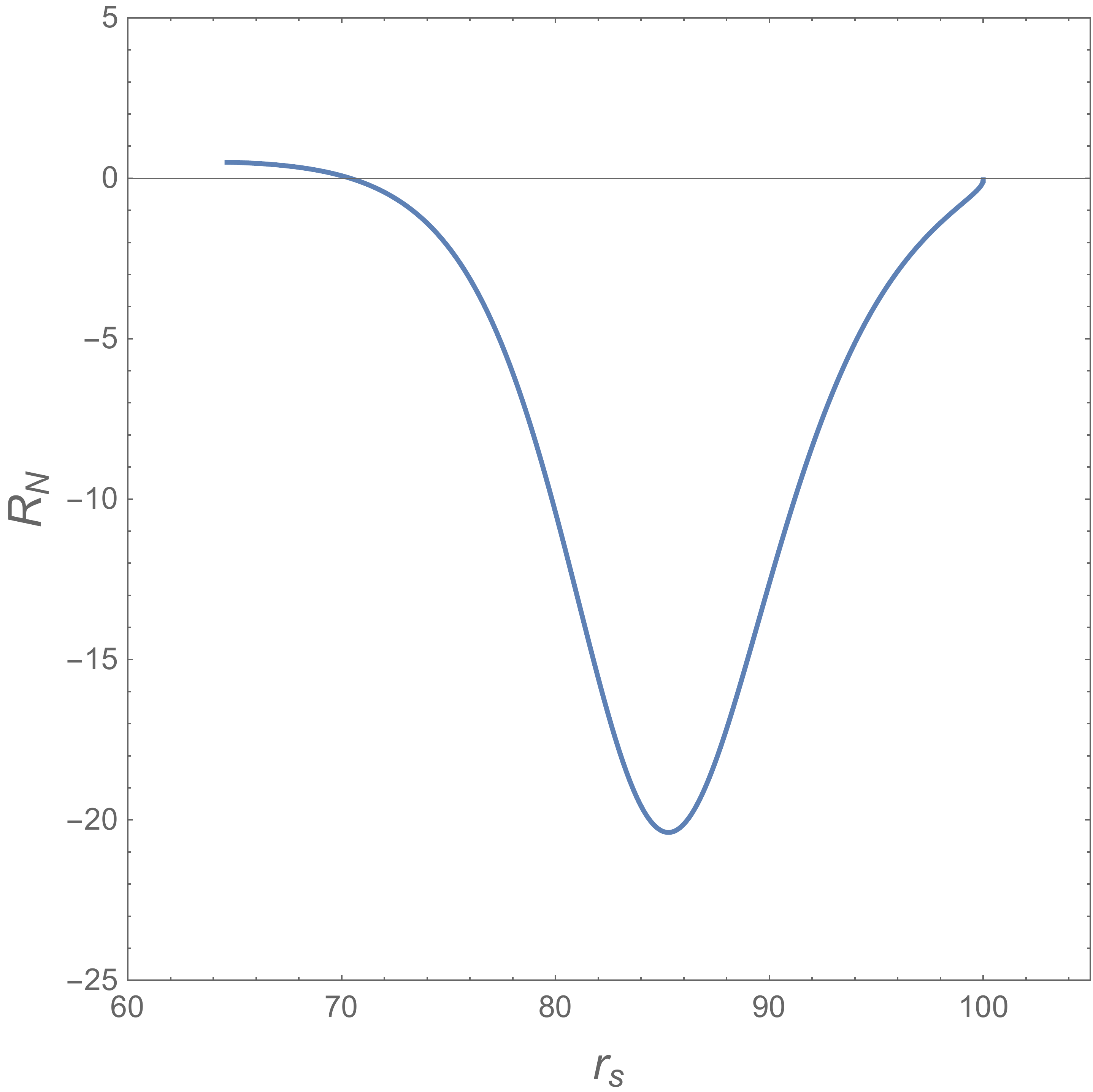}
	          	}
	          	\caption{The variation of $R_N$ in terms of the shadow radius $r_s$ with $P = 1.5$ and $r_o=100$. }\label{Fig8}
	          \end{figure}
    \item $P = 1.5$ in Fig.\ref{Fig8}:
              There is no divergent point and the black hole is in the supercritical phase. The normalized curvature scalar $R_N$ decreases first and then increases at this time.
\end{itemize}
By comparing with that in \cite{Wei:2019yvs}, our results suggest that observation of black hole shadows indeed give correct information about the small/large black hole phase structure and its underlying microstructure.

\section{\textsf{Hawking-Page phase structure and microstructure}} \label{III}

In this section, we would like to study the Hawking-Page (HP) phase transition of the RN-AdS black hole with shadow formalism. The Hawking temperature Eq.(\ref{temperature}) in terms of the thermodynamic quantities $S$, $P$ and $\Phi$ is
\begin{equation}
	T=\frac{6-\Phi^2+3PS}{8 \sqrt{S}},     \label{TemperatureS}
\end{equation}
Solving the entropy from the Hawking temperature above, it yields
\begin{align}
	S= \frac{ 32 T^2 - 3P \left(6-\Phi^2 \right) \pm  8T \sqrt{16 T^2 -3 P  \left(6-\Phi^2 \right)}}{9 P^2},     \label{entropy2}
\end{align}
which implies there are two branches of $S$.

Then we turn our attention to HP phase transition. The Hawking mechanism tells us that a black hole can establish thermal equilibrium with the environment through the exchange of matter and energy. Due to the conservation of charge, a black hole with a fixed value of $Q$ won't undergo the HP phase transition to thermal vacuum which is electrically neutral. So that we will study the HP phase transition of the charged AdS black hole in the grand canonical ensemble with fixed electric potential $\Phi$. By regarding the mass as enthalpy, the Gibbs free energy has the form of
\begin{align}
	G(T,P,\Phi) &=M-TS - \Phi Q= \frac{1}{4} \sqrt{ S } \left( 6-\Phi^2 - PS \right).    \label{Gibbs} 
\end{align}
The entropy Eq.(\ref{entropy2}) implied that the Gibbs free energy also has two branches. The particle number of the thermal gas in AdS spacetime will vary with temperature, and the Gibbs free energy of thermal AdS background is zero. Meanwhile, the free energy of a black hole is non-zero, so the critical condition for the HP phase transition is the vanishing point of the Gibbs free energy
\begin{equation}
	6-\Phi^2 - PS=0,    \label{HPcritical}
\end{equation}
which gives the HP temperature $T_{\mathrm{HP}}$ as
\begin{equation}
	T_{\mathrm{HP}} = \frac{\sqrt{ P \left( 6-\Phi^2 \right) }}{2}.    \label{HPtemperature}
\end{equation}
The potential is constraint by $|\Phi| \leq \sqrt{6}$ as the Hawking temperature of a black hole is positive.

By substituting Eq.(\ref{entropy2}) into Eq.(\ref{Gibbs}), we illustrate the Gibbs free energy curve as function of $T$ in Fig.\ref{Fig9}. 
\begin{figure}[!h]
	\centering
	\includegraphics[width=8cm]{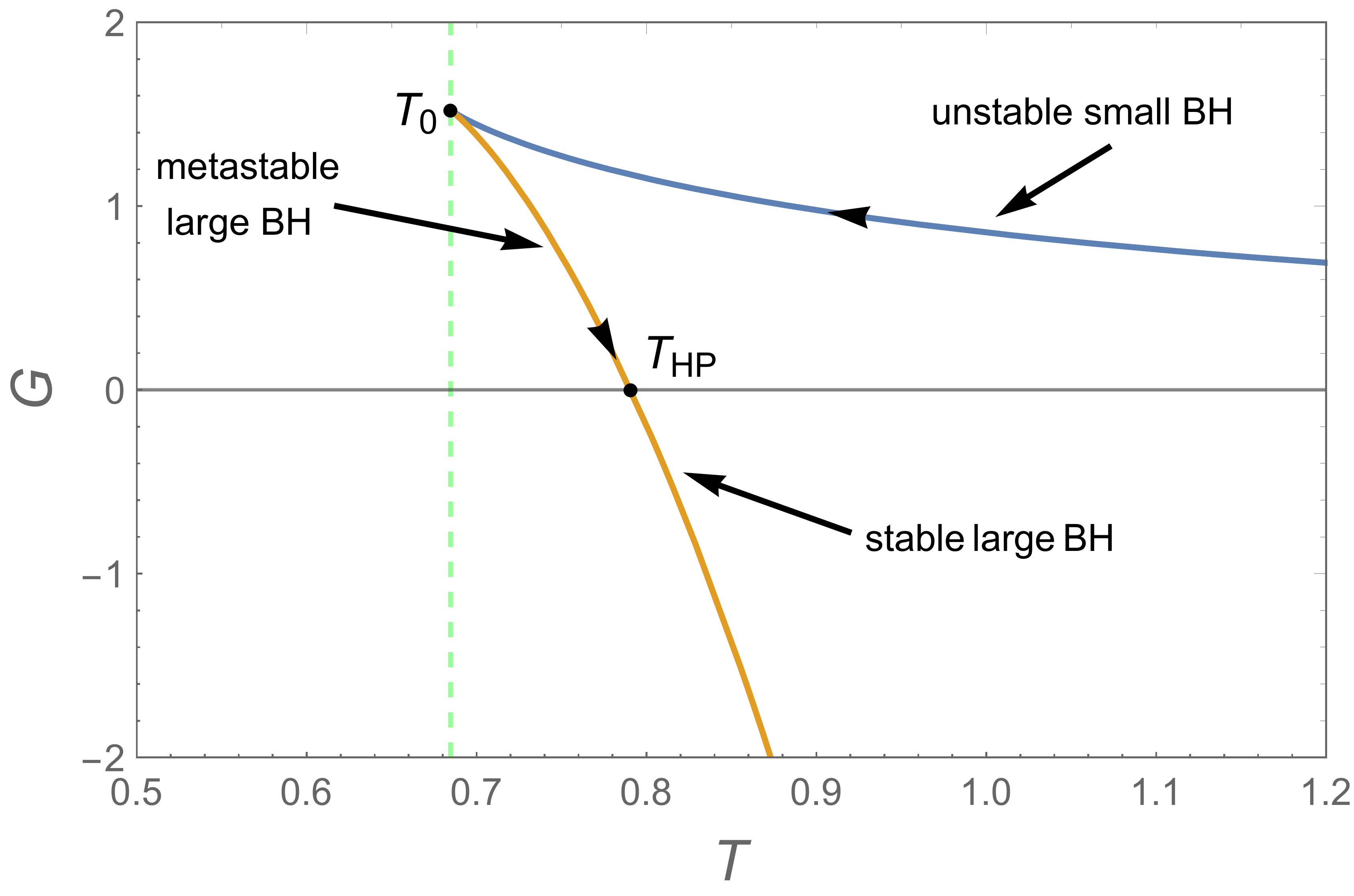}
	\caption{The Gibbs free energies of RN-AdS black hole. $T_\mathrm{HP}$ and $T_0$ denote the Hawking-Page phase transition temperature and minimum temperature. The arrow indicates the increased direction of the black hole event horizon radius.}\label{Fig9}
\end{figure}
The blue and orange curves represent the unstable small and large black holes, respectively. The black arrow along the curve indicates the direction in which the radius of the event horizon $r_h$ increases. There exists a minimum temperature $T_0$ coming from the intersection of the two branches of $G$
\begin{equation}
	T_0=\frac{ \sqrt{3 P \left( 6-\Phi^2 \right)} }{4}.    \label{T0}
\end{equation}
Below this temperature, the background is the thermal radiation. Increasing the temperature, the thermal radiation have the probability to form a large black hole, which is represented by the part between $T_0$ and $T_{\mathrm{HP}}$ in the orange solid curve in Fig.\ref{Fig9}. The HP phase transition occurs at $T_{\mathrm{HP}}$, above which it is preferred to form a stable large black hole.

To explore the relation between Gibbs free energy and the shadow radius with reduced parameters, we build the silhouette graph from the e-index of Gibbs free energy in Fig.(\ref{Fig10}). 
\begin{figure}[!h]
	\centering
	\subfigure[]{
		\includegraphics[width=6.4cm]{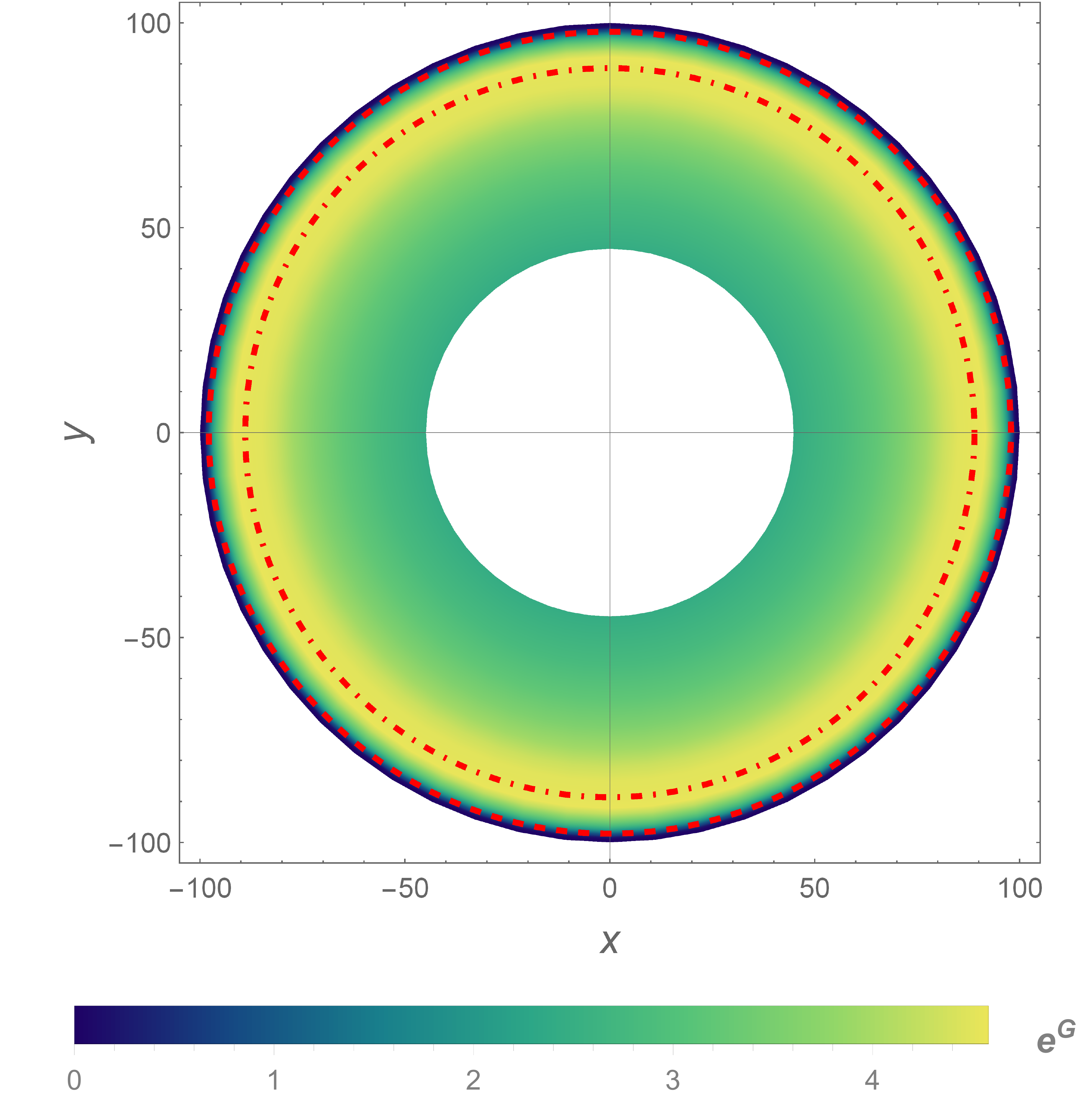}
	}
	\quad
	\subfigure[]{
		\includegraphics[width=6.4cm]{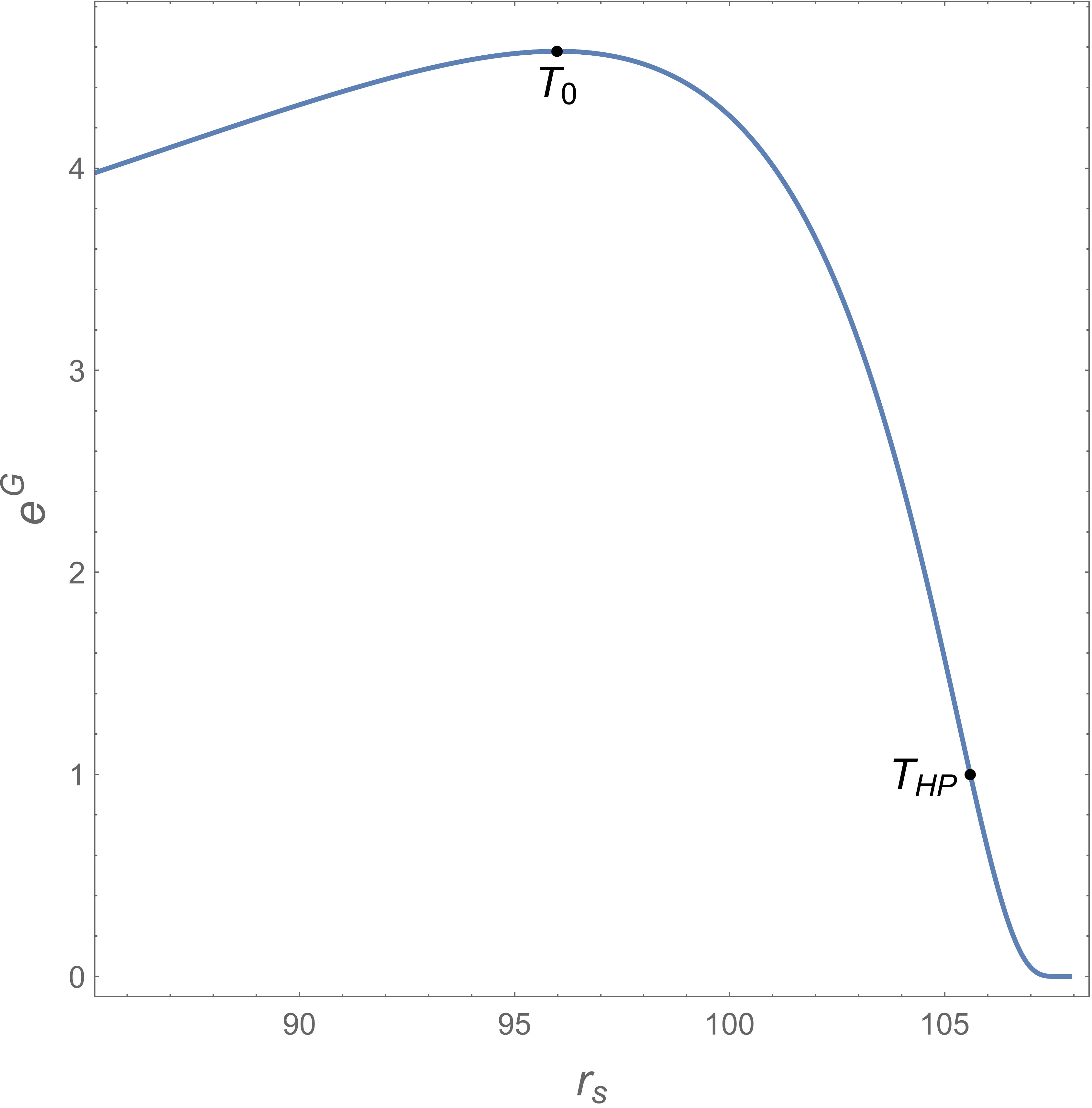}
	}
	\caption{The variation of $e^G$ in terms of shadow radius $r_s$, here we have set $P = 0.5$ and $\Phi=1$. The blue solid curve in both figure represent $G=0$ and $T_\mathrm{HP}$, respectively.}\label{Fig10}
\end{figure}
The right panel is the $e^{G}$ curve. There are two special points in the figure: $r_{s0}$ and $r_\mathrm{HP}$ corresponding to the minimum temperature $T_0$ and the Hawking-Page temperature $T_\mathrm{HP}$, respectively. The shadow of the black hole smaller than $r_{s0}$ corresponds to the unstable phase, which indicates it is more preferable to stay in the thermal radiation phase. The middle segment between $r_{s0}$ and $r_{\mathrm{HP}}$ represent the metastable phase, it is possible to form a large black hole. When the radius of shadow is larger than $r_{\mathrm{HP}}$, the AdS background thermal radiation undergoes HP phase transition to form a black hole. We depict the silhouette of black hole shadow in Fig.\ref{Fig11}(a). The legend-bar corresponds to the value of $e^{G}$. The red dot-dashed and dashed circle correspond to $r_{s0}$ and $r_{\mathrm{HP}}$, respectively.  

After the above discussion, we would like to know the underlying microstructure of the black hole on the HP phase transition curve in Fig.\ref{Fig9}. Substituting Eq.(\ref{TemperatureS}) and Eq.(\ref{HPcritical}) into Eq.(\ref{sc2}), the normalized curvature scalar along the HP curve in reduced $(T,V)$ space can be obtained as
\begin{equation}
	R_N= \frac{\left(6-\Phi^2 \right) \left( -8 T \sqrt{ S}+ 6- \Phi^2 \right)}{2 \left(-4 T \sqrt{S} + 6- \Phi^2 \right)^2}.    \label{scHP}
\end{equation}
To visualize the underlying microstructure, we depict the silhouette of shadow with the normalized curvature scalar $R_N$ in Fin.\ref{Fig11}.
\begin{figure}[!h]
	\centering
	\subfigure[]{
		\includegraphics[width=6.5cm]{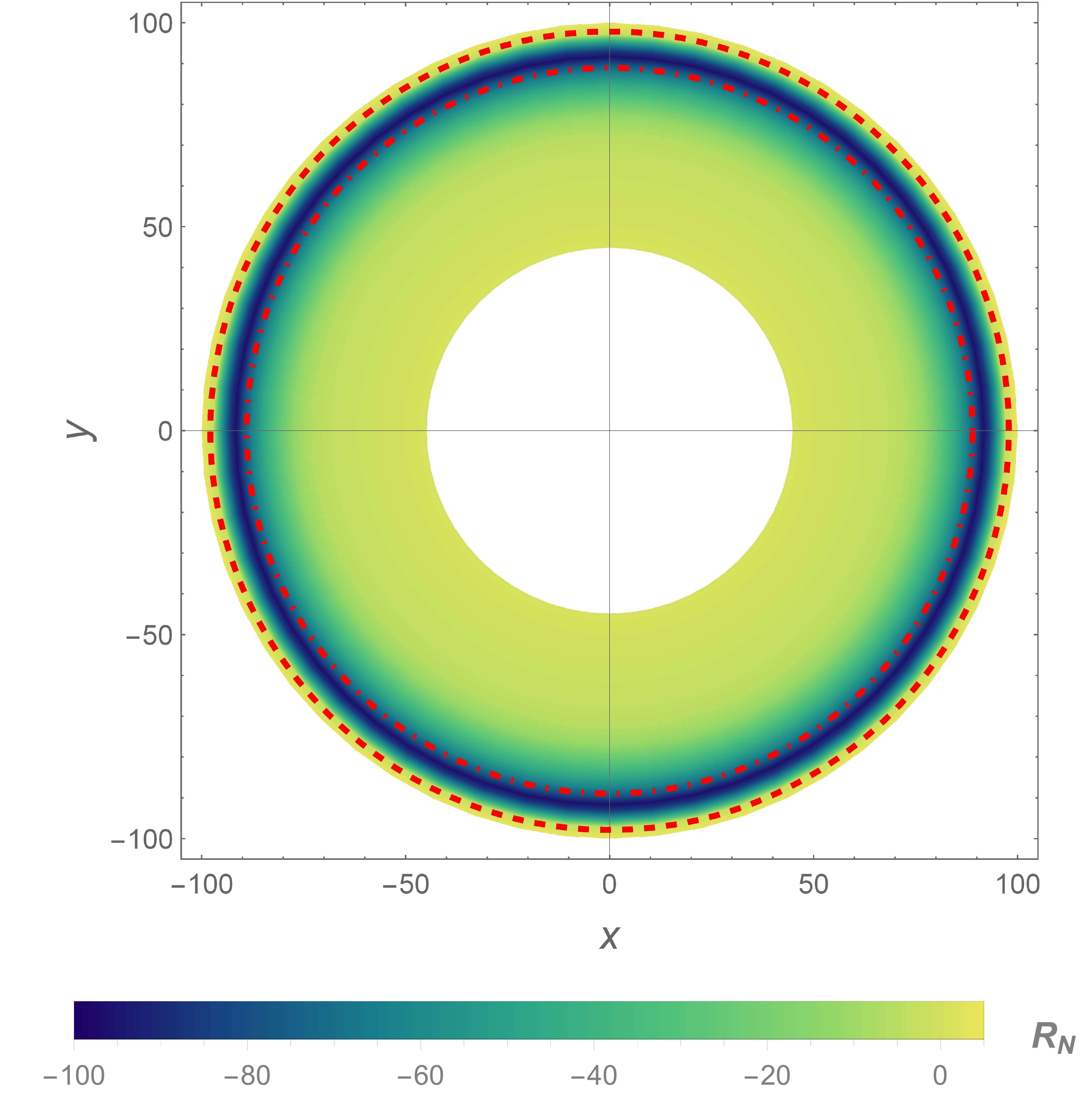}
	}
	\quad
	\subfigure[]{
		\includegraphics[width=7cm]{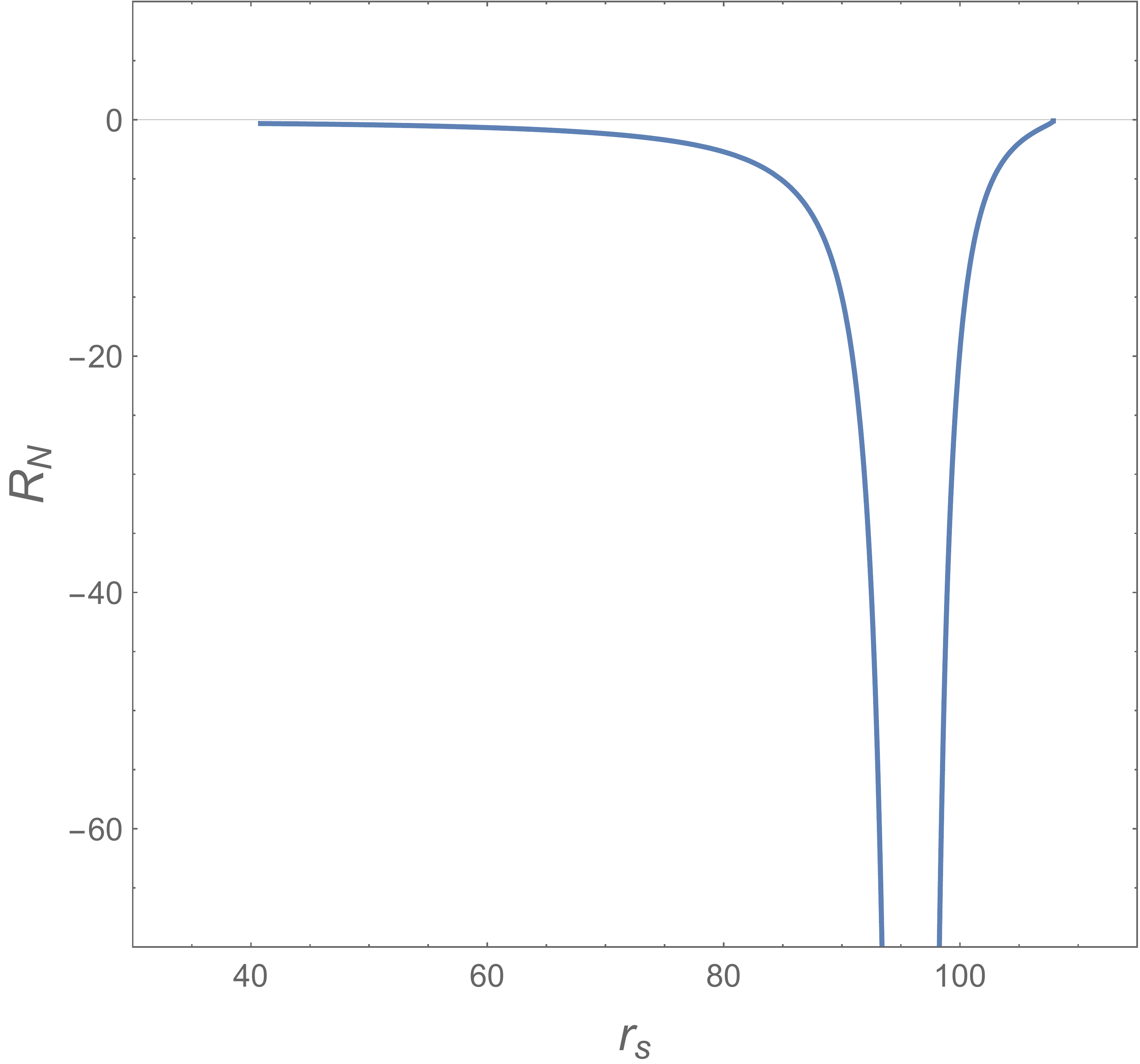}
	}
	\caption{The variation of $R_N$ in terms of the shadow radius $r_s$, here we have set $P = 0.5$ and $\Phi=1$.}\label{Fig11}
\end{figure}
The red dot-dashed and dashed circle correspond to $T_0$ and $T_\mathrm{HP}$, respectively. The thermodynamic phase of a black hole is divided into three parts by these two red circles. They are the unstable small black hole, the metastable large black hole and the stable large black hole with the shadow radius increasing. As shown by the legend-bar, the curvature scalar $R_N$ is non-monotonic. The black hole always have a negative value of $R_N$ implying the interaction between the fluctuate state is attractive domain. As discussed above, the information about the Hawking-Page phase transition and the underlying microstructure of the corresponding black hole can be directly observed through its shadow.

By substituting Eq.(\ref{HPtemperature}) into Eq.(\ref{scHP}), we find that the normalized curvature scalar of the critical point of HP phase transition of the RN-AdS black hole is  also a constant
\begin{equation}
	R_{N\mathrm{HP}}= -\frac{3}{2},    \label{RNHP}
\end{equation}
which is consistent with the result reported in \cite{Wei:2020kra}. It is implied that the value of the normalized curvature scalar at the HP phase transition point is a threshold as that the repulsive interaction of the thermal radiation turns to the attractive interaction inside the black hole at the critical point.

\section{\textsf{Conclusion}}

In this paper, we investigate the phase structure and the underlying microstructure of the RN-AdS black hole using shadow formalism. After regarding the negative cosmological constant as the thermodynamic pressure, the black hole exhibits abundant thermodynamic phase structure. We hope to get information about the thermodynamics of the black hole from the observation of its shadow. For that purpose, we first revisit the radius of the shadow of the RN-AdS black hole and plot it with respect to the event horizon radius in Fig.\ref{Fig2}. As shown in the figure, the curve of the shadow radius is divided into three parts by two points $a$ and $b$: the curve of the shadow radius on the left of point $a$ is cast off since it will lead to a negative temperature, while the part to the right of point $b$ can not be adopted for the observer is always assumed to be located out of the photon sphere. But don't be frustrated, as in Fig.\ref{Fig3} we find that the phase structure of the RN-AdS black hole can be completely reflected and directly ``observed'' from its shadow within this parameter range.

Furthermore, we associate respectively the Van der Waals-like and Hawking-Page phase transition with the black hole shadow in the extended phase space. It can be seen in Fig.\ref{Fig3}(b) that the phase structure shows the same behavior as that in Fig.\ref{Fig3}(a), which indicates that the shadow can give the correct information about the small/large phase transition. This conclusion also holds for the Hawking-Page phase transition after analyzing its phase structure in Fig.\ref{Fig9} and Fig.\ref{Fig10}. And then we turn our attention to the underlying microstructure described by the Rupperner geometry with thermodynamic line element originated from the internal energy of black holes. Since the thermodynamic metric is degenerated in the case of the vanishing heat capacity at constant volume, the normalized curvature scalar $R_N$ is introduced. By associating $R_N$ with the shadow radius, we build the silhouette of shadow in a series of figures. In these figures, circles with different radii correspond to the shadows of black holes in different states, and the variation of $R_N$ is revealed within the color gradient of the legend-bar. The result shows that the shadow of a charged static spherically symmetric black hole can correctly reflect its underlying microstructure. What's more, the normalized curvature scalar at the critical point of HP phase transition of the RN-AdS black hole has a constant value of $-3/2$, which supports the conjecture proposed in \cite{Wei:2020kra}. The close connection between black hole dynamics and its thermodynamics opens a new window for us to understand black holes from the perspective of observation, and we will pay more attention to it in the future.

\section*{\textsf{Acknowledgment}}
This work is supported by the financial supports from the National Natural Science Foundation of China (Grant Nos.12047502, 11947208), the China Postdoctoral Science Foundation (Grant Nos.2017M623219, 2020M673460), Basic Research Program of Natural Science of Shaanxi Province (Grant No.2019JQ-081) and Double First-Class University Construction Project of Northwest University.
\end{spacing}

\providecommand{\href}[2]{#2}\begingroup
\footnotesize\itemsep=0pt
\providecommand{\eprint}[2][]{\href{http://arxiv.org/abs/#2}{arXiv:#2}}
\providecommand{\doi}[2]{\href{http://dx.doi.org/#2}{#1}}

\endgroup


\begin{thebibliography}{99}
	
\bibitem{Akiyama:2019cqa}
K.~Akiyama \textit{et al.} [Event Horizon Telescope],
\emph{First M87 Event Horizon Telescope Results. I. The Shadow of the Supermassive Black Hole},
\href{https://iopscience.iop.org/article/10.3847/2041-8213/ab0ec7}{\emph{Astrophys. J. Lett. }{\bfseries 875}, L1 (2019)},
[\href{https://arxiv.org/abs/1906.11238}{{\ttfamily arXiv:1906.11238}}].

\bibitem{Akiyama:2019bqs}
K.~Akiyama \textit{et al.} [Event Horizon Telescope],
\emph{First M87 Event Horizon Telescope Results. IV. Imaging the Central Supermassive Black Hole},
\href{https://iopscience.iop.org/article/10.3847/2041-8213/ab0e85}{\emph{Astrophys. J. Lett. }{\bfseries 875}, L4 (2019)},
[\href{https://arxiv.org/abs/1906.11241}{{\ttfamily arXiv:1906.11241}}].

\bibitem{Abbott:2016blz}
B.~P.~Abbott \textit{et al.} [LIGO Scientific and Virgo],
\emph{Observation of Gravitational Waves from a Binary Black Hole Merger},
\href{https://journals.aps.org/prl/abstract/10.1103/PhysRevLett.116.061102}{\emph{Phys. Rev. Lett. }{\bfseries 116}, 061102 (2016)},
[\href{https://arxiv.org/abs/1602.03837}{{\ttfamily arXiv:1602.03837}}].

\bibitem{Tsukamoto:2014tja}
N.~Tsukamoto, Z.~Li and C.~Bambi,
\emph{Constraining the spin and the deformation parameters from the black hole shadow},
\href{https://doi.org/10.1088/1475-7516/2014/06/043}{\emph{JCAP }{\bfseries 06}, 043 (2014)},
[\href{https://arxiv.org/abs/1403.0371}{{\ttfamily arXiv:1403.0371}}].

\bibitem{Zakharov:2011zz}
A.~F.~Zakharov, F.~De Paolis, G.~Ingrosso and A.~A.~Nucita,
\emph{Shadows as a tool to evaluate black hole parameters and a dimension of spacetime},
\href{https://doi.org/10.1016/j.newar.2011.09.002}{\emph{New Astron. Rev. }{\bfseries 56}, 64-73 (2012)}.

\bibitem{Kumar:2018ple}
R.~Kumar and S.~G.~Ghosh,
\emph{Black Hole Parameter Estimation from Its Shadow},
\href{https://iopscience.iop.org/article/10.3847/1538-4357/ab77b0}{\emph{Astrophys. J. }{\bfseries 892}, 78 (2020)},
[\href{https://arxiv.org/abs/1811.01260}{{\ttfamily arXiv:1811.01260}}].

\bibitem{Allahyari:2019jqz}
A.~Allahyari, M.~Khodadi, S.~Vagnozzi and D.~F.~Mota,
\emph{Magnetically charged black holes from non-linear electrodynamics and the Event Horizon Telescope},
\href{https://iopscience.iop.org/article/10.1088/1475-7516/2020/02/003}{\emph{JCAP }{\bfseries 02}, 003 (2020)},
[\href{https://arxiv.org/abs/1912.08231}{{\ttfamily arXiv:1912.08231}}].

\bibitem{Khodadi:2020jij}
M.~Khodadi, A.~Allahyari, S.~Vagnozzi and D.~F.~Mota,
\emph{Black holes with scalar hair in light of the Event Horizon Telescope},
\href{https://iopscience.iop.org/article/10.1088/1475-7516/2020/09/026}{\emph{JCAP }{\bfseries 09}, 026 (2020)},
[\href{https://arxiv.org/abs/2005.05992}{{\ttfamily arXiv:2005.05992}}].

\bibitem{Xu:2018mkl}
Z.~Xu, X.~Hou and J.~Wang,
\emph{Possibility of Identifying Matter around Rotating Black Hole with Black Hole Shadow},
\href{https://doi.org/10.1088/1475-7516/2018/10/046}{\emph{JCAP }{\bfseries 10}, 046 (2018)},
[\href{https://arxiv.org/abs/1806.09415}{{\ttfamily arXiv:1806.09415}}].

\bibitem{Saurabh:2020zqg}
K.~Saurabh and K.~Jusufi,
\emph{Imprints of dark matter on black hole shadows using spherical accretions},
\href{https://doi.org/10.1140/epjc/s10052-021-09280-9}{\emph{Eur. Phys. J. }{\bfseries C81}, 490 (2021)},
[\href{https://arxiv.org/abs/2009.10599}{{\ttfamily arXiv:2009.10599}}].

\bibitem{Synge:1966}
J.~L.~Synge,
\emph{The Escape of Photons from Gravitationally Intense Stars},
\href{https://doi.org/10.1093/mnras/131.3.463}{\emph{Mon. Not. Roy. Astron. Soc. }{\bfseries 131}, 463-466 (1966)}.

\bibitem{Vries:2000}
A.~ de Vries,
\emph{The apparent shape of a rotating charged black hole, closed photon orbits and the bifurcation set $A_4$},
\href{https://doi.org/10.1088/0264-9381/17/1/309}{\emph{Class. Quant. Grav. }{\bfseries 17}, 123 (2000)}.

\bibitem{Hioki:2008zw}
K.~Hioki and U.~Miyamoto,
\emph{Hidden symmetries, null geodesics, and photon capture in the Sen black hole},
\href{https://link.aps.org/doi/10.1103/PhysRevD.78.044007}{\emph{Phys. Rev. }{\bfseries D78}, 044007 (2008)},
[\href{https://arxiv.org/abs/0805.3146}{{\ttfamily arXiv:0805.3146}}].

\bibitem{Eiroa:2017uuq}
E.~F.~Eiroa and C.~M.~Sendra,
\emph{Shadow cast by rotating braneworld black holes with a cosmological constant},
\href{https://doi.org/10.1140/epjc/s10052-018-5586-6}{\emph{Eur. Phys. J. }{\bfseries C78}, 91 (2018)},
[\href{https://arxiv.org/abs/1711.08380}{{\ttfamily arXiv:1711.08380}}].

\bibitem{Bambi:2008jg}
C.~Bambi and K.~Freese,
\emph{Apparent shape of super-spinning black holes},
\href{https://doi.org/10.1103/PhysRevD.79.043002}{\emph{Phys. Rev. }{\bfseries D79}, 043002 (2009)},
[\href{https://arxiv.org/abs/0812.1328}{{\ttfamily arXiv:0812.1328}}].

\bibitem{Bambi:2010hf}
C.~Bambi and N.~Yoshida,
\emph{Shape and position of the shadow in the $\delta = 2$ Tomimatsu-Sato space-time},
\href{https://doi.org/10.1088/0264-9381/27/20/205006}{\emph{Class. Quant. Grav. }{\bfseries 27}, 205006 (2010)},
[\href{https://arxiv.org/abs/1004.3149}{{\ttfamily arXiv:1004.3149}}].

\bibitem{Wang:2017qhh}
M.~Wang, S.~Chen and J.~Jing,
\emph{Shadows of Bonnor black dihole by chaotic lensing},
\href{https://doi.org/10.1103/PhysRevD.97.064029}{\emph{Phys. Rev. }{\bfseries D97}, 064029 (2018)},
[\href{https://arxiv.org/abs/1710.07172}{{\ttfamily arXiv:1710.07172}}].

\bibitem{Guo:2018kis}
M.~Guo, N.~A.~Obers and H.~Yan,
\emph{Observational signatures of near-extremal Kerr-like black holes in a modified gravity theory at the Event Horizon Telescope},
\href{https://doi.org/10.1103/PhysRevD.98.084063}{\emph{Phys. Rev. }{\bfseries D98}, 084063 (2018)},
[\href{https://arxiv.org/abs/1806.05249}{{\ttfamily arXiv:1806.05249}}].

\bibitem{Yan:2019etp}
H.~Yan,
\emph{Influence of a plasma on the observational signature of a high-spin Kerr black hole},
\href{https://doi.org/10.1103/PhysRevD.99.084050}{\emph{Phys. Rev. }{\bfseries D98}, 084050 (2019)},
[\href{https://arxiv.org/abs/1903.04382}{{\ttfamily arXiv:1903.04382}}].

\bibitem{Konoplya:2019sns}
R.~A.~Konoplya,
\emph{Shadow of a black hole surrounded by dark matter},
\href{https://doi.org/10.1016/j.physletb.2019.05.043}{\emph{Phys. Lett. }{\bfseries B795}, 1-6 (2019)},
[\href{https://arxiv.org/abs/1905.00064}{{\ttfamily arXiv:1905.00064}}].

\bibitem{Cunha:2018acu}
P.~V.~P.~Cunha and C.~A.~R.~Herdeiro,
\emph{Shadows and strong gravitational lensing: a brief review},
\href{https://doi.org/10.1007/s10714-018-2361-9}{\emph{Gen. Rel. Grav. }{\bfseries 50}, 42 (2018)},
[\href{https://arxiv.org/abs/1801.00860} {{\ttfamily arXiv:1801.00860}}].

\bibitem{Bekenstein:1973ur}
J.~D. Bekenstein, 
\emph{Black holes and entropy},
\href{https://doi.org/10.1103/PhysRevD.7.2333}{\emph{Phys. Rev. } {\bfseries D7}, 2333--2346 (1973)}.

\bibitem{Hawking:1974sw}
S.~W. Hawking, 
\emph{Particle Creation by Black Holes},
\href{https://doi.org/10.1007/BF02345020, 10.1007/BF01608497}{\emph{Commun. Math. Phys.} {\bfseries 43}, 199--220 (1975)}.

\bibitem{Bardeen:1973gs}
J.~M. Bardeen, B.~Carter and S.~W. Hawking, 
\emph{The Four laws of black hole mechanics}, 
\href{https://doi.org/10.1007/BF01645742}{\emph{Commun. Math. Phys. } {\bfseries 31}, 161--170 (1973)}.

\bibitem{Hawking:1976de}
S.~W. Hawking, 
\emph{Black Holes and Thermodynamics},
\href{https://doi.org/10.1103/PhysRevD.13.191}{\emph{Phys. Rev. } {\bfseries D13}, 191--197 (1976)}.

\bibitem{Hawking:1982dh}
S.~W. Hawking and D.~N. Page, 
\emph{Thermodynamics of Black Holes in anti-DeSitter Space}, 
\href{https://doi.org/10.1007/BF01208266}{\emph{Commun. Math. Phys. } {\bfseries 87}, (1983) 577}.

\bibitem{Witten:1998zw}
E.~Witten, \emph{Anti-de Sitter space, thermal phase transition, and confinementin gauge theories},
\href{https://doi.org/10.4310/ATMP.1998.v2.n3.a3}{\emph{Adv. Theor. Math. Phys. } {\bfseries 2}, (1998) 505--532},
[\href{https://arxiv.org/abs/hep-th/9803131}{{\ttfamily arXiv:hep-th/9803131}}].

\bibitem{Kastor:2009wy}
D.~Kastor, S.~Ray and J.~Traschen, 
\emph{Enthalpy and the Mechanics of AdS Black Holes}, 
\href{https://doi.org/10.1088/0264-9381/26/19/195011}{\emph{Class. Quant. Grav. } {\bfseries 26}, 195011 (2009)},
[\href{https://arxiv.org/abs/0904.2765}{{\ttfamily arXiv:0904.2765}}].

\bibitem{Dolan:2011xt}
B.~P. Dolan, 
\emph{Pressure and volume in the first law of black hole thermodynamics},
\href{https://doi.org/10.1088/0264-9381/28/23/235017}{\emph{Class. Quant. Grav. } {\bfseries 28}, 235017 (2011)},
[\href{https://arxiv.org/abs/1106.6260}{{\ttfamily arXiv:1106.6260}}].

\bibitem{Cvetic:2010jb}
M.~Cvetic, G.~W. Gibbons, D.~Kubiznak and C.~N. Pope, 
\emph{Black Hole Enthalpy and an Entropy Inequality for the Thermodynamic Volume},
\href{https://doi.org/10.1103/PhysRevD.84.024037}{\emph{Phys. Rev. }{\bfseries D84}, 024037 (2011)},
[\href{https://arxiv.org/abs/1012.2888}{{\ttfamily arXiv:1012.2888}}].

\bibitem{Kubiznak:2012wp}
D.~Kubiznak and R.~B. Mann, 
\emph{P-V criticality of charged AdS black holes},
\href{https://doi.org/10.1007/JHEP07(2012)033}{\emph{JHEP} {\bfseries 07}, 033 (2012)}, 
[\href{https://arxiv.org/abs/1205.0559}{{\ttfamily arXiv:1205.0559}}].

\bibitem{Kubiznak:2016qmn}
D.~Kubiznak, R.~B. Mann and M.~Teo, 
\emph{Black hole chemistry: thermodynamics with Lambda}, \href{https://doi.org/10.1088/1361-6382/aa5c69}{\emph{Class. Quant. Grav.} {\bfseries 34}, 063001 (2017)},
[\href{https://arxiv.org/abs/1608.06147}{{\ttfamily arXiv:1608.06147}}].

 \bibitem{Toledo:2019amt}
J.~M. Toledo, and V.~B. Bezerra,
\emph{Some remarks on the thermodynamics of charged AdS black holes with cloud of strings and quintessence},
\href{https://doi.org/10.1140/epjc/s10052-019-6616-8}{\emph{Eur. Phys. J. }{\bfseries C79}, 110 (2019)}.

\bibitem{Hendi:2012um}
S.~H. Hendi and M.~H. Vahidinia, 
\emph{Extended phase space thermodynamics and P-V criticality of black holes with a nonlinear source},
\href{https://doi.org/10.1103/PhysRevD.88.084045}{\emph{Phys. Rev. }{\bfseries D88}, 084045 (2013)},
[\href{https://arxiv.org/abs/1212.6128}{{\ttfamily arXiv:1212.6128}}].

\bibitem{Wei:2012ui}
S.-W. Wei and Y.-X. Liu, 
\emph{Critical phenomena and thermodynamic geometry of charged Gauss-Bonnet AdS black holes},
\href{https://doi.org/10.1103/PhysRevD.87.044014}{\emph{Phys. Rev. }{\bfseries D87}, 044014 (2013)},
[\href{https://arxiv.org/abs/1209.1707}{{\ttfamily arXiv:1209.1707}}].

\bibitem{Cai:2013qga}
R.-G. Cai, L.-M. Cao, L.~Li and R.-Q. Yang, 
\emph{P-V criticality in the extended phase space of Gauss-Bonnet black holes in AdS space},
\href{https://doi.org/10.1007/JHEP09(2013)005}{\emph{JHEP } {\bfseries 09}, 005 (2013)}, 
[\href{https://arxiv.org/abs/1306.6233}{{\ttfamily arXiv:1306.6233}}].

\bibitem{Zhao:2013oza}
R.~Zhao, H.-H. Zhao, M.-S. Ma and L.-C. Zhang, 
\emph{On the critical phenomena and thermodynamics of charged topological dilaton AdS black holes},
\href{https://doi.org/10.1140/epjc/s10052-013-2645-x}{\emph{Eur. Phys. J. }{\bfseries C73}, 2645  (2013)},
[\href{https://arxiv.org/abs/1305.3725}{{\ttfamily arXiv:1305.3725}}].

\bibitem{Mo:2013ela}
J.-X. Mo and W.-B. Liu, 
\emph{Ehrenfest scheme for P-V criticality in the extended phase space of black holes},
\href{https://doi.org/10.1016/j.physletb.2013.10.045}{\emph{Phys. Lett. }{\bfseries B727}, 336--339 (2013)}.

\bibitem{Altamirano:2013ane}
N.~Altamirano, D.~Kubiznak and R.~B. Mann, 
\emph{Reentrant phase transitions in rotating anti-de Sitter black holes},
\href{https://doi.org/10.1103/PhysRevD.88.101502}{\emph{Phys. Rev. }{\bfseries D88}, 101502 (2013)},
[\href{https://arxiv.org/abs/1306.5756}{{\ttfamily arXiv:1306.5756}}].

\bibitem{Spallucci:2013osa}
E.~Spallucci and A.~Smailagic, 
\emph{Maxwell's equal area law for charged Anti-de Sitter black holes},
\href{https://doi.org/10.1016/j.physletb.2013.05.038}{\emph{Phys. Lett. }{\bfseries B723}, 436--441  (2013)},
[\href{https://arxiv.org/abs/1305.3379}{{\ttfamily arXiv:1305.3379}}].

\bibitem{Xu:2014tja}
H.~Xu, W.~Xu and L.~Zhao, 
\emph{Extended phase space thermodynamics for third order Lovelock black holes in diverse dimensions},
\href{https://doi.org/10.1140/epjc/s10052-014-3074-1}{\emph{Eur. Phys. J. }{\bfseries C74}, 3074 (2014)},
[\href{https://arxiv.org/abs/1405.4143}{{\ttfamily arXiv:1405.4143}}].

\bibitem{Miao:2018fke}
Y.-G. Miao and Z.-M. Xu, 
\emph{Parametric phase transition for a Gauss-Bonnet AdS black hole}, 
\href{https://doi.org/10.1103/PhysRevD.98.084051}{\emph{Phys. Rev. } {\bfseries D98}, 084051 (2018)},
[\href{https://arxiv.org/abs/1806.10393}{{\ttfamily arXiv:1806.10393}}].

\bibitem{Miao:2016ulg}
Y.-G. Miao and Z.-M. Xu, 
\emph{Phase transition and entropy inequality of noncommutative black holes in a new extended phase space},
\href{https://doi.org/10.1088/1475-7516/2017/03/046}{\emph{JCAP } {\bfseries 1703}, 046 (2017)}, 
[\href{https://arxiv.org/abs/1604.03229}{{\ttfamily arXiv:1604.03229}}].

\bibitem{Xu:2013zea}
W.~Xu, H.~Xu and L.~Zhao, 
\emph{Gauss-Bonnet coupling constant as a free thermodynamical variable and the associated criticality},
\href{https://doi.org/10.1140/epjc/s10052-014-2970-8}{\emph{Eur. Phys. J. }{\bfseries C74}, 2970 (2014)},
[\href{https://arxiv.org/abs/1311.3053}{{\ttfamily arXiv:1311.3053}}].

\bibitem{Wei:2017mwc}
S.~W.~Wei and Y.~X.~Liu,
\emph{Photon orbits and thermodynamic phase transition of $d$-dimensional charged AdS black holes},
\href{https://doi.org/10.1103/PhysRevD.97.104027}{\emph{Phys. Rev. }{\bfseries D97}, 104027 (2018)},
[\href{https://arxiv.org/abs/1711.01522}{{\ttfamily arXiv:1711.01522}}].

\bibitem{Wei:2018aqm}
S.~W.~Wei, Y.~X.~Liu and Y.~Q.~Wang,
\emph{Probing the relationship between the null geodesics and thermodynamic phase transition for rotating Kerr-AdS black holes},
\href{https://doi.org/10.1103/PhysRevD.99.044013}{\emph{Phys. Rev. }{\bfseries D99}, 044013 (2019)},
[\href{https://arxiv.org/abs/1807.03455}{{\ttfamily arXiv:1807.03455}}].

\bibitem{Zhang:2019tzi}
M.~Zhang, S.~Z.~Han, J.~Jiang and W.~B.~Liu,
\emph{Circular orbit of a test particle and phase transition of a black hole},
\href{https://doi.org/10.1103/PhysRevD.99.065016}{\emph{Phys. Rev. }{\bfseries D99}, 065016 (2019)},
[\href{https://arxiv.org/abs/1903.08293}{{\ttfamily arXiv:1903.08293}}].

\bibitem{Zhang:2019glo}
M.~Zhang and M.~Guo,
\emph{Can shadows reflect phase structures of black holes?}
\href{https://doi.org/10.1140/epjc/s10052-020-8389-5}{\emph{Eur. Phys. J. }{\bfseries C80}, 790 (2020)},
[\href{https://arxiv.org/abs/1909.07033}{{\ttfamily arXiv:1909.07033}}].

\bibitem{Belhaj:2020nqy}
A.~Belhaj, L.~Chakhchi, H.~El Moumni, J.~Khalloufi and K.~Masmar,
\emph{Thermal Image and Phase Transitions of Charged AdS Black Holes using Shadow Analysis},
\href{https://doi.org/10.1142/S0217751X20501705}{\emph{Int. J. Mod. Phys. }{\bfseries A35}, 2050170 (2020)},
[\href{https://arxiv.org/abs/2005.05893}{{\ttfamily arXiv:2005.05893}}].

\bibitem{PhysRevA.24.488}
G.~Ruppeiner, 
\emph{Application of Riemannian geometry to the thermodynamics of a simple fluctuating magnetic system},
\href{https://doi.org/10.1103/PhysRevA.24.488}{\emph{Phys. Rev. } {\bfseries A24}, 488--492 (1981)}.

\bibitem{Ruppeiner:1983zz}
G.~Ruppeiner, 
\emph{Thermodynamic Critical Fluctuation Theory?}
\href{https://doi.org/10.1103/PhysRevLett.50.287}{\emph{Phys. Rev. Lett. }{\bfseries 50}, 287--290 (1983)}.

\bibitem{Ruppeiner:1995zz}
G.~Ruppeiner, 
\emph{Riemannian geometry in thermodynamic fluctuation theory},
\href{https://doi.org/10.1103/RevModPhys.67.605}{\emph{Rev. Mod. Phys. }{\bfseries 67}, 605--659 (1995)}.

\bibitem{Ruppeiner:2008kd}
G.~Ruppeiner,
\emph{Thermodynamic curvature and phase transitions in Kerr-Newman black holes},
\href{https://doi.org/10.1103/PhysRevD.78.024016}{\emph{Phys. Rev. }{\bfseries D78}, 024016 (2008)},
[\href{https://arxiv.org/abs/0802.1326}{{\ttfamily arXiv:0802.1326}}].

\bibitem{Ruppeiner:2010}
G.~Ruppeiner,
\emph{Thermodynamic curvature measures interactions},
\href{http://dx.doi.org/10.1119/1.3459936}{\emph{Am. J. Phys. }{\bfseries 78}, 1107 (2010)},
[\href{https://arxiv.org/abs/1007.2160}{{\ttfamily arXiv:1007.2160}}].

\bibitem{Sahay:2010wi}
A.~Sahay, T.~Sarkar and G.~Sengupta,
\emph{Thermodynamic Geometry and Phase Transitions in Kerr-Newman-AdS Black Holes},
\href{https://doi.org/10.1007/JHEP04(2010)118}{\emph{JHEP }{\bfseries 04}, 118 (2010)},
[\href{https://arxiv.org/abs/1002.2538}{{\ttfamily arXiv:1002.2538}}].

\bibitem{Zhang:2015ova}
J.~L.~Zhang, R.~G.~Cai and H.~Yu,
\emph{Phase transition and thermodynamical geometry of Reissner-Nordstr\"om-AdS black holes in extended phase space},
\href{https://doi.org/10.1103/PhysRevD.91.044028}{\emph{Phys. Rev. }{\bfseries D91}, 044028 (2015)},
[\href{https://arxiv.org/abs/1502.01428}{{\ttfamily arXiv:1502.01428}}].

\bibitem{Wei:2017icx}
S.-W. Wei, B.~Liang and Y.-X. Liu, 
\emph{Critical phenomena and chemical potential of a charged AdS black hole},
\href{https://doi.org/10.1103/PhysRevD.96.124018}{\emph{Phys. Rev. }{\bfseries D96}, 124018 (2017)},
[\href{https://arxiv.org/abs/1705.08596}{{\ttfamily arXiv:1705.08596}}].

\bibitem{Chaturvedi:2017vgq}
P.~Chaturvedi, S.~Mondal and G.~Sengupta, 
\emph{Thermodynamic Geometry of Black Holes in the Canonical Ensemble},
\href{https://doi.org/10.1103/PhysRevD.98.086016}{\emph{Phys. Rev. }{\bfseries D98}, 086016 (2018)},
[\href{https://arxiv.org/abs/1705.05002}{{\ttfamily arXiv:1705.05002}}].

\bibitem{Xu:2019gqm}
Z.~M.~Xu, B.~Wu and W.~L.~Yang,
\emph{Ruppeiner thermodynamic geometry for the Schwarzschild-AdS black hole},
\href{https://doi.org/10.1103/PhysRevD.101.024018}{\emph{Phys. Rev. D }{\bfseries 101}, 024018 (2020)},
[\href{https://arxiv.org/abs/1910.12182}{{\ttfamily arXiv:1910.12182}}].

\bibitem{Wei:2019yvs}
S.~W.~Wei, Y.~X.~Liu and R.~B.~Mann,
\emph{Ruppeiner Geometry, Phase Transitions, and the Microstructure of Charged AdS Black Holes},
\href{https://doi.org/10.1103/PhysRevD.100.124033}{\emph{Phys. Rev. D }{\bfseries 100}, 124033 (2019)},
[\href{https://arxiv.org/abs/1909.03887}{{\ttfamily arXiv:1909.03887}}].

\bibitem{Xu:2020ftx}
Z.~M.~Xu, B.~Wu and W.~L.~Yang,
\emph{Diagnosis inspired by the thermodynamic geometry for different thermodynamic schemes of the charged BTZ black hole},
\href{https://doi.org/10.1140/epjc/s10052-020-08563-x}{\emph{Eur. Phys. J. }{\bfseries C80}, 997 (2020)},
[\href{https://arxiv.org/abs/2002.00117}{{\ttfamily arXiv:2002.00117}}].

\bibitem{Wei:2019uqg}
S.-W. Wei, Y.-X. Liu and R.~B. Mann, 
\emph{Repulsive Interactions and Universal Properties of Charged Anti-de Sitter Black Hole Microstructures},
\href{https://doi.org/10.1103/PhysRevLett.123.071103}{\emph{Phys. Rev. Lett. }{\bfseries 123}, 071103 (2019)},
[\href{https://arxiv.org/abs/1906.10840}{{\ttfamily arXiv:1906.10840}}].

\bibitem{Xu:2020gud}
Z.~M.~Xu, B.~Wu and W.~L.~Yang,
\emph{Ruppeiner thermodynamic geometry for the Schwarzschild-AdS black hole},
\href{https://doi.org/10.1103/PhysRevD.101.024018}{\emph{Phys. Rev. }{\bfseries D101}, 024018 (2020)},
[\href{https://arxiv.org/abs/1910.12182}{{\ttfamily arXiv:1910.12182}}].

\bibitem{Cai:2021fpr}
X.~C.~Cai and Y.~G.~Miao,
\emph{Can shadows connect black hole microstructures?}
[\href{https://arxiv.org/abs/2101.10780}{{\ttfamily arXiv:2101.10780}}].

\bibitem{Cai:2021uov}
X.~C.~Cai and Y.~G.~Miao,
\emph{Can we know about black hole thermodynamics through shadows?}
[\href{https://arxiv.org/abs/2107.08352}{{\ttfamily arXiv:2107.08352}}].


\bibitem{Perlick:2015vta}
V.~Perlick, O.~Y.~Tsupko and G.~S.~Bisnovatyi-Kogan,
\emph{Influence of a plasma on the shadow of a spherically symmetric black hole},
\href{https://doi.org/10.1103/PhysRevD.92.104031}{\emph{Phys. Rev. }{\bfseries D92}, 104031 (2015)},
[\href{https://arxiv.org/abs/1507.04217}{{\ttfamily arXiv:1507.04217}}].

\bibitem{Perlick:2018iye}
V.~Perlick, O.~Y.~Tsupko and G.~S.~Bisnovatyi-Kogan, 
\emph{Black hole shadow in an expanding universe with a cosmological constant},
\href{https://link.aps.org/doi/10.1103/PhysRevD.97.104062}{\emph{Phys. Rev. }{\bfseries D97}, 104062 (2018)},
[\href{https://arxiv.org/abs/1804.04898}{{\ttfamily arXiv:1804.04898}}].

\bibitem{Wu:2020fij}
B.~Wu, C.~Wang, Z.~M.~Xu and W.~L.~Yang,
\emph{Ruppeiner geometry and thermodynamic phase transition of the black hole in massive gravity},
[\href{https://arxiv.org/abs/2006.09021}{{\ttfamily arXiv:2006.09021}}].

\bibitem{Wei:2020kra}
S.~W.~Wei, Y.~X.~Liu and R.~B.~Mann,
\emph{Novel dual relation and constant in Hawking-Page phase transitions},
\href{https://doi.org/10.1103/PhysRevD.102.104011}{\emph{Phys. Rev. }{\bfseries D102}, 104011 (2020)},
[\href{https://arxiv.org/abs/2006.11503}{{\ttfamily arXiv:2006.11503}}].

\bibitem{Yan:2021uzw}
D.~W.~Yan, Z.~R.~Huang and N.~Li,
\emph{Hawking-Page phase transitions of charged AdS black holes surrounded by quintessence},
\href{https://doi.org/10.1088/1674-1137/abc0cf}{\emph{Chin. Phys. }{\bfseries C45}, 015104 (2021)}.



\end{thebibliography}
\end{document}